\documentclass[11pt,a4paper]{article}
\usepackage[T1]{fontenc}
\usepackage{lmodern}
\usepackage[margin=1in]{geometry}
\usepackage{amsmath,amssymb}
\usepackage{graphicx,subcaption}
\usepackage{float,enumitem}
\floatstyle{ruled}
\newfloat{algorithm}{tbp}{loa}
\floatname{algorithm}{Algorithm}
\usepackage[numbers,sort&compress]{natbib}
\usepackage[hidelinks]{hyperref}
\numberwithin{equation}{section}

\newcommand{\probDist}{\mu}
\newcommand{\deterministicModel}{f}
\newcommand{\R}{\mathbb{R}}
\newcommand{\nOutputDimensions}{L}
\newcommand{\inputVectorRandom}{X}
\newcommand{\nvar}{d}
\newcommand{\indOutputs}{\ell}
\newcommand{\indPoints}{i}
\newcommand{\nDoE}{n}
\newcommand{\outputPoint}{y}
\newcommand{\outputHighDimensional}{\outputDeterministic_{\indOutputs}}
\newcommand{\nbasis}{n_{b}}
\newcommand{\inputPoint}{x}
\newcommand{\coef}{c}
\newcommand{\indBasis}{q}
\newcommand{\DoE}{\mathcal{X}}
\newcommand{\dataMatrixPCA}{\mathcal{Y}}
\newcommand{\basis}{v}
\newcommand{\centeredData}{\widetilde{\dataMatrixPCA}}
\newcommand{\covarianceMatrixPCA}{\Sigma}
\newcommand{\coefficientsMatrixPCA}{C}
\newcommand{\basisMatrixPCA}{V}
\newcommand{\dimensionMean}{\overline{\outputPoint}}
\newcommand{\meanGPR}{m}
\newcommand{\kernelGPR}{K}
\newcommand{\GaussianProcess}{Z}
\newcommand{\MeanFunction}{m}
\newcommand{\KernelFunction}{K}
\newcommand{\condGP}{\GaussianProcess_n}
\newcommand{\ConditionalMean}{\MeanFunction_c}
\newcommand{\ConditionalKernel}{\KernelFunction_c}
\newcommand{\KernelMatrix}{\KernelFunction(\DoE, \DoE)}
\newcommand{\KernelVector}[1]{\KernelFunction(\DoE, #1)}
\newcommand{\KernelVectorRow}[1]{\KernelFunction(#1, \DoE)}
\newcommand{\deterministicModelGPR}{g}
\newcommand{\SobolNormalized}[1]{S_{#1}}
\newcommand{\SobolUnnormalized}[1]{D_{#1}}
\newcommand{\sobolTotD}[1]{\SobolUnnormalized{#1}^{\text{T}}}
\newcommand{\sobolClosed}[1]{\SobolNormalized{#1}^{\text{c}}}
\newcommand{\sobolClosedD}[1]{\SobolUnnormalized{#1}^{\text{c}}}
\newcommand{\totalVariance}{D}
\newcommand{\GSI}{\mathrm{GSI}^{\text{c}}}
\newcommand{\inputPointRandom}[1]{\inputVectorRandom_{#1}}
\newcommand{\Var}{\mathbb{V}\mathrm{ar}}
\newcommand{\indexSet}{I}
\newcommand{\exceptIndexSet}{- \indexSet}
\newcommand{\Esp}{\mathbb{E}}
\newcommand{\dimInt}{\{1, \dots, d\}}
\newcommand{\Cov}{\mathbb{C}\mathrm{ov}}
\newcommand{\pfEstim}[1]{{\widehat{#1}}^{\mathrm{pf}}}
\newcommand{\basisVec}{\basis_{., \indOutputs}}
\newcommand{\domain}{\mathbb{X}}
\newcommand{\Tr}[1]{\mathrm{Tr}\left[ #1 \right]}
\newcommand{\inputVector}{\ensuremath{X} }
\newcommand{\DoEpoint}[1]{x^{(#1)}}
\newcommand{\DoEset}{\{\DoEpoint{1}, \dots, \DoEpoint{\nDoE} \}}
\newcommand{\outputDeterministic}{y}
\newcommand{\initialheight}{H}
\newcommand{\plasticviscosity}{\mu_B}
\newcommand{\density}{\rho}
\newcommand{\yieldstress}{\tau_c}
\newcommand{\liftingvelocity}{V_L}
\newcommand{\runout}{x_f}
\newcommand{\binghamnumber}{Bi}
\newcommand{\inertianumber}{I}
\newcommand{\uniformDist}{\mathcal{U}}
\newcommand{\runoutsteady}{x_{f_{\infty}}}
\newcommand{\lengthreservoir}{l}
\newcommand{\flowrate}{q}
\newcommand{\sobollifting}{S_{\liftingvelocity}}
\newcommand{\sobolviscosity}{S_{\plasticviscosity}}
\newcommand{\sobolinitial}{S_{\initialheight}}
\newcommand{\sobolyield}{S_{\yieldstress}}
\newcommand{\soboldensity}{S_{\density}}

\title{Global sensitivity analysis of transient regimes in non-Newtonian dam-break flows applied in rheometry}
\author{Yuri Taglieri Sáo, Geraldo de Freitas Maciel, João Batista Pereira}
\date{November 18, 2026}

\begin{document}
\maketitle
\begin{abstract}
Non-Newtonian dam-break flows are a classical problem with numerous engineering applications. These flows are particularly important in alternative rheometry, where they are used to rapidly and affordably estimate rheological properties, such as in the slump test and Bostwick consistometer test. However, the lifting dynamics of the containing obstacle introduce inertial effects that can compromise the accurate estimation of these properties. These effects are influenced by the lifting velocity, the fluid’s viscosity, and the initial geometry of the sample. This study aims to quantify the influence of lifting velocity and to provide physical insights into how inertial effects influence the flow. A global sensitivity analysis (GSA) framework adapted for functional outputs (e.g., time series) is employed, using data generated from high-fidelity simulations with ANSYS Fluent. The GSA produces \textit{sensitivity curves} that track the evolving importance of each input parameter over time. Three distinct flow regimes are identified, namely mobilization, transition and viscous regimes, each governed by different dominant factors. The first two regimes are significantly affected by the lifting velocity, revealing the influence of inertial effects, whereas the viscous regime is dominated by fluid rheology and shows a increasing importance of the yield stress. These results are contextualized in terms of traditional rheometry tests.

\end{abstract}

\section{Introduction}

Non-Newtonian dam-break flows are a classical problem with applications in geophysics, hydraulics, and materials science \citep{Ancey2007, Ancey2009, Coussot2017, Gildeh2021, Tan2017, Bouvet2010, Pereira2021}. It is considered a valuable first-approximation model for landslides and dam-breach flows \citep{Jeyapalan1983, Ancey2012, Muchiri2024a}, and is also widely used for benchmarking numerical solvers \citep{Li2018, Muchiri2024b, BrondaniMinussi2025}. At both laboratory and prototype scales, this problem has a remarkable importance in alternative rheometry, which focuses on fast and low-cost estimation of the rheological properties of flowing materials \citep{Pashias1996, Clayton2003, Roussel2005a, Balmforth2007, Gao2015a, Chaparian2018}. We refer the reader to \citet{Coussot2005} and \citet{Coussot2014} for further details on alternative rheometry techniques.

Within this context, several experimental procedures rely on steady-state (or arrested-state) measurements, such as the vertical and/or horizontal displacement of the fluid’s free surface, commonly referred to as slump and spread (or runout), respectively. The arrested regime is characterized by a balance between the pressure gradient and the fluid’s intrinsic viscoplastic properties, such as the yield stress. Empirical correlations have been developed to link experimental measurements of slump or spread to yield stress values. Examples of such tests include the slump test \citep{Pashias1996, Clayton2003}, mini-slump test \citep{Gao2015a, Tan2017}, and L-box test \citep{Chaparian2018}. Other procedures investigate the transient phase of dam-break flows, capturing the evolution of slump or spread over time to focus on the viscous properties of the fluid. For instance, the Bostwick consistometer test \citep{Perona2005, Balmforth2007} uses a reference numerical solution of a partial differential equation describing shallow and slow Herschel–Bulkley flows. By measuring the time required for the wavefront to reach predefined positions, this test enables estimation of the consistency index, flow index, and yield stress. Other methods, such as the modified slump test \citep{Ferraris1998} and the automated slump test \citep{Chidiac2000, Pereira2021, Pereira2024}, also explore the transient phase and allow the extraction of viscous parameters.

However, these rheological estimations assume ideal experimental conditions, which may not hold in practical, in-field applications \citep{Pereira2021}. Recent studies have demonstrated that the lifting velocity of the mould or gate influences both the transient and steady-state behaviors of slump and spread in dam-break experiments \citep{Gao2015a, Zheng2024}, although other works do not identify significative effect over the slump measure \citep{Pashias1996, Zhang2024}. The lifting mechanism introduces inertial effects into the flow \citep{Roussel2005a, Gao2015a, Zhang2024}, making the application of empirical and quasi-static models less reliable. These inertial effects can depend on several factors, including the fluid's viscosity, the initial height of the sample, and the lifting velocity \citep{Roussel2005a, Gao2015a, Valette2021}. Although some studies have investigated the impact of lifting dynamics and associated inertia, they have typically been restricted to a few representative scenarios \citep{Gao2015a, Zheng2024}. While these studies provide valuable localized insights, a global evaluation of parameter influence on non-Newtonian dam-break flows remains largely unexplored in the literature.

To address this gap, the present study aims to quantify the importance of uncertain input variables over the time development of the flow runout, by considering a gradual dam-break of non-Newtonian fluid flow over an idealized geometry. More specifically, a probabilistic framework of uncertainty quantification and global sensitivity analysis is applied, which enables the estimation of sensitivity indices representing the influence of rheological parameters, initial conditions and, specially, lifting velocity. Furthermore, the time evolution of sensitivity indices allows to perform physical interpretations of the runout development regarding viscous and inertial effects of the flow. This framework relies on a vast range of scenarios generated by a high-fidelity computational fluid dynamics (CFD) solver. Sensitivity indices are computed over the entire time domain, yielding sensitivity curves that illustrate how the relative importance of each parameter evolves as the flow progresses. These indices support a physically grounded interpretation of parameter influence at any point in time. By analyzing both the sensitivity curves and the corresponding numerical data, three distinct flow regimes were identified, each characterized by different dominant parameters. Finally, these findings are discussed in the context of traditional alternative rheometry methods, with specific reference to the slump test and the Bostwick consistometer test.

The work is organized as follows. Section \ref{sec:dam_break_flows} presents a brief review on prototypical-scale dam-break flows of non-Newtonian fluids. Section \ref{sec:methodology} details the methodologies of this work, separated in a deterministic methodology for data acquisition in Subsec. \ref{subsec:numerical} and in a probabilistic framework for statistical treatment of data in Subsec. \ref{subsec:UQ} and \ref{subsec:GSA_chap3}. Then, Section \ref{sec:results_chap3} shows the results and discussion and Section \ref{sec:conclusion_chap3} closes the work with the conclusions.

\section{Prototypical dam-break flows of non-Newtonian fluids}
\label{sec:dam_break_flows}

In this section, we review important works that explore the dam-break flows of non-Newtonian fluids in prototypical scale in the alternative rheometry context. Initial efforts to understand the behavior of viscoplastic materials under dam-break flows were focused on steady-state configurations, i.e. the arrested state of the flow. In this context based on the final configuration of the fluid sample, we highlight planar-geometry tests, such as the inclined channel \citep{Coussot1995, Lajeunesse2004} and L-box \citep{Nguyen2006, Chaparian2018} tests, and the axisymmetric-geometry tests, e.g. the slump-test \citep{Pashias1996, Clayton2003, Roussel2005b, Pereira2022} and the mini-slump test \citep{Roussel2005a, Bouvet2010, Gao2015a}. Planar-geometry tests commonly employ the final wavefront position as characteristic length, known as \textit{runout} or \textit{spread}, whereas slump-tests can use both the spread and the vertical displacement of the free-surface, or simply \textit{slump} measure. The slump measure usually reaches the final configuration in a small time-scale \citep{Pashias1996}, usually in the order of few seconds, while the spread measure tends to develop from minutes up to hours \citep{Coussot1995}. In general, these tests, known as stoppage tests, show a good capability to estimate the yield stress of the materials, usually relating a characteristic length with dimensionless forms of yield stress scaled by the initial sample height and density \citep{Pashias1996, Clayton2003, Saak2004, Pereira2022}.   

While the arrested state is limited to the estimation of yield stress, the transient state of these tests has been explored to estimate viscous properties. A well-known technique is the Bostwick consistometer test \citep{Perona2005, Balmforth2007, DellaVecchia2019, Birnbaum2021}, which monitors the time taken by the wavefront to reach three monitors in the channel. These values of time are compared with numerical solutions of governing equations based on lubrication theory; then, rheological parameters can be estimated. The modified slump-test \citep{Ferraris1998, Chidiac2000, Pereira2024} and mini-slump test \citep{Choi2016, Guo2024} follow the same logic: the time period for the free-surface to reach a specific height is used to estimate empirically the viscosity. A particular limitation is reported by \citet{Castruccio2010}, which is the difficulty of controlling the levels of strain or applied stress, governed by parameters such as volume, initial height and density.

In general, the aforementioned studies indicate that the measured properties of these types of dam-break flows are affected by initial geometry (e.g. initial height) and properties of the fluid (density, yield stress and viscous parameters, for instance). However, both the steady and transient-state tests of dam-break problems are prone to experimental uncertainties. One of the relevant sources of uncertainty is the lifting procedure of the cone in slump tests or gate in channels, more specifically the lifting velocity of the obstacle. \citet{Gao2015a} argues that lifting velocity has a major influence in the slump measure in mini-slump tests for low yield stress and viscosity, whereas the spread is less affected. This is corroborated by \citet{Zheng2024}, which identified that faster lifting velocities lead to smaller calculated yield stress. The reason is that a fast lifting introduces significant inertial effects in comparison with slow lifting, which affects the final slump and, at a lower extent, the final spread. On the other hand, other works indicate no significant influence of lifting velocity to the yield stress measurement in the arrested state \citep{Pashias1996, Zhang2024}. In the transient scenario, the lifting velocity may modify the behavior of the slump and spread over time, as slower lifting velocities slows down the release of fluid and, consequently, its development. 

The lifting velocity appears to be an important factor for experiments, since it seems to modify the measurable properties of the dam-break flows. Most of the investigations analyze isolated/local scenarios and do not quantify parameter sensitivity over time, specially when dealing with lifting velocity. This gap motivates the application of a rigorous global sensitivity analysis framework, which is capable of quantifying the influence of input parameters over the output data, e.g. evolution of spread over time. The background of this framework is introduced in the next section.

\section{Methodology}
\label{sec:methodology}

In this work, we employ a combination of uncertainty quantification (UQ) approach and global sensitivity analysis (GSA) techniques. The joint usage of UQ and GSA is very convenient, since UQ propagates and quantifies the input uncertainty over outputs and provides statistical measures (mean, variance, confidence intervals, etc), while GSA provides indicators to explain aspects of outputs uncertainty, e.g. important and unimportant input variables and possible interactions between them. Here, we opt for a non-intrusive strategy through the Monte Carlo method, where a deterministic model acts as a black-box code, receiving uncertain inputs scenarios and generating quantities of interest (QoI) without any modifications of the computational code. The deterministic model used to simulate the dam-break flows of non-Newtonian fluids is described in subsection \ref{subsec:numerical}. Next, the UQ approach is detailed in subsection \ref{subsec:UQ} and two issues inherent to the uncertainty propagation must be taken into account. Firstly, the QoI evaluated in this work (runout over time) is functional and, consequently, high-dimensional; to address this issue, a dimensionality-reduction technique is required. Secondly, the deterministic code is computationally expensive and its direct use turns the probabilistic analysis of data prohibitive; to solve this issue, a common approach is to build a statistical model (metamodel or surrogate model) with relatively few points, being capable of producing accurate outputs with significantly less computational effort. Both dimensionality-reduction and metamodeling techniques allows to perform GSA efficiently and the specific techniques are detailed in subsection \ref{subsec:GSA_chap3}.

\subsection{Deterministic model}
\label{subsec:numerical}

\begin{figure}[H] 
\centering
\includegraphics[width=0.7\textwidth]{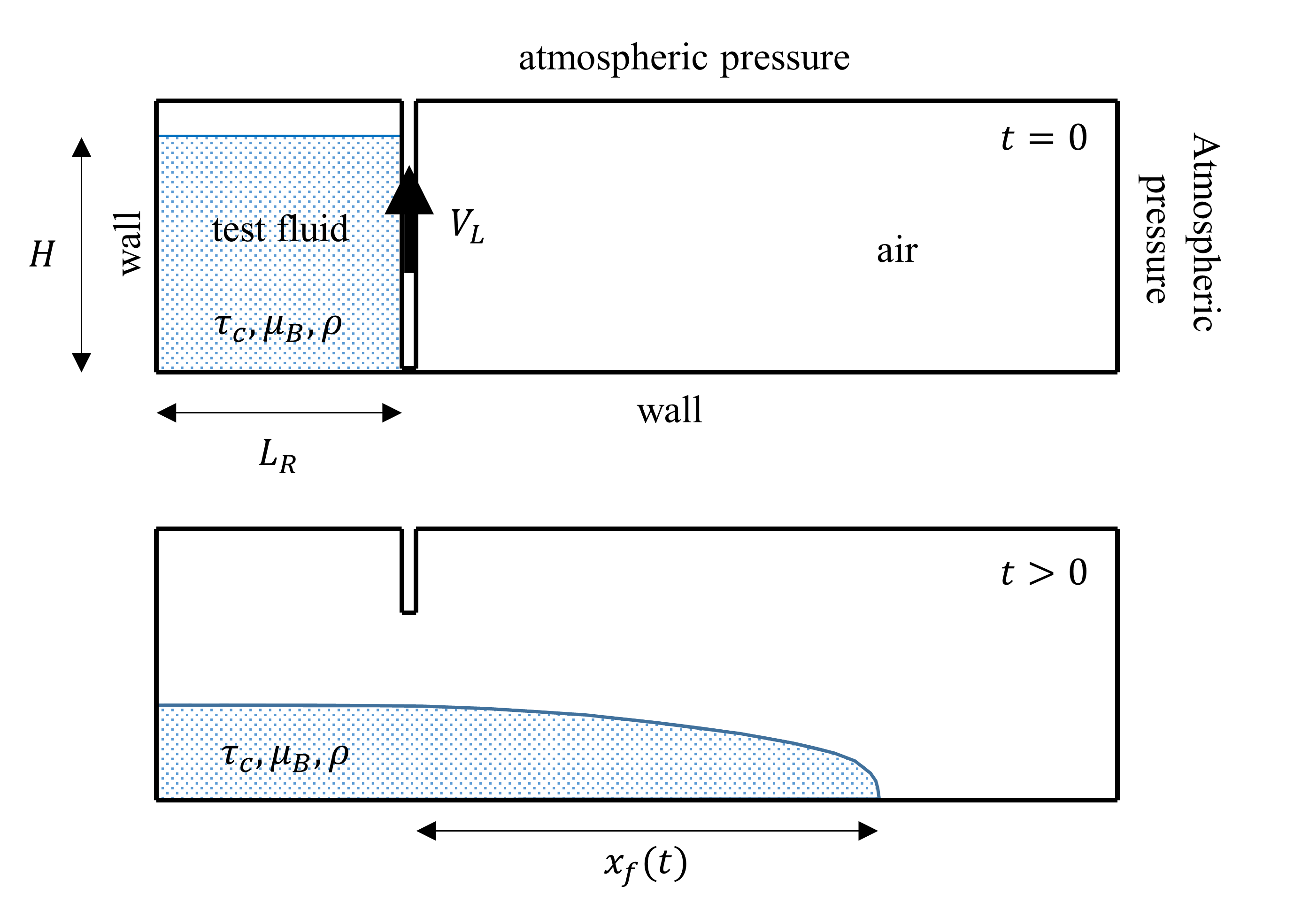}
\caption{Schematic of the idealized gradual dam-break case with boundary conditions and input variables.}\label{Fig.schematic_chap3}
\end{figure}

An idealized gradual dam-break flow consists of a known volume of material ($Vol=0.28 \ m^3$), with initial height $\initialheight$, inside a reservoir delimited by the walls and a gate, which is lifted with a finite velocity $\liftingvelocity$ and the material flows downstream a horizontal plane or channel (Fig. \ref{Fig.schematic_chap3}). Here, we consider fluid modeled rheologically by the Bingham model (Eq. \ref{eq:bingham_model}), which presents yield stress $\yieldstress$ and plastic viscosity $\plasticviscosity$ with density $\density$. After the release of the fluid, the position of the wavefront $\runout(t)$ is captured over time; therefore, this is the quantity of interest of the analysis presented herein.

Data obtained in this work was obtained by 2D numerical simulations using ANSYS Fluent r20.0, which employs the Finite Volume Method to discretize and solve the system of equations of continuity and momentum from fluid mechanics \citep{Matsson2022}. Figure \ref{Fig.schematic_chap3} shows the computational domain and boundary conditions. Structured quadrangular meshes with $\Delta x = 5 \times 10^{-4}$ $m$ of resolution were used and the Layering technique was applied to address the moving boundary (lifting gate) \citep{Gao2015a, Pereira2022}. The grid size uncertainty related to this resolution varied from $0.96 \%$ to $1.51 \%$ according to the grid convergence index (GCI) method. An adaptive scheme of time discretization in function of the Courant-Friedrichs-Lewy (CFL) condition was employed; we employed $\text{CFL}=0.25$, with uncertainties estimated in the order of $0.1 \%$ to $0.6 \%$. The estimation of grid size and time discretization uncertainties are shown with more details in \citet{Sao2021}. The number of iterations per time step varied according to the convergence criteria: convergence was satisfied if the governing equations residuals were smaller than $10^{-6}$ or if no significant variation of residuals were detected in the last $20$ iterations. The multiphasic model Volume of Fluid (VoF) was used to model different phases by introducing the variable Volume Fraction $\alpha_{(\cdot)}$ ($\alpha_1=0$ for air, $\alpha_2=1$ for test fluid, where $\alpha_1 + \alpha_2 = 1$) and to track the free surface of the test fluid ($0 < \alpha_2 < 1$) \citep{Gopala2008}. To model the non-Newtonian fluid rheology, the regularized Bingham model was employed \citep{Bird1983, Balmforth2014}, whose constitutive equation in apparent viscosity $\eta$ form is given by Eq. \ref{eq:bingham_model}. 

\begin{equation}
\label{eq:bingham_model}
    \eta = 
    \begin{cases}
        \frac{\tau_c}{\dot{\gamma}_c} \left( 2 - \frac{\dot{\gamma}}{\dot{\gamma}_c}\right) + \mu_B & \text{if } \dot{\gamma} \leq \dot{\gamma}_c \\
        \frac{\tau_c}{\dot{\gamma}} + \mu_B & \text{if } \dot{\gamma} > \dot{\gamma}_c
    \end{cases}
\end{equation}

where $\dot{\gamma}$ is the shear rate and $\dot{\gamma}_c$ is the critical shear rate. We considered $\dot{\gamma}_c = 10^{-2}$ $s^{-1}$, which is sufficient to regularize the viscosity \citep{Gao2015a, Pereira2022}. 

In order to acquire the runout distance over time, a longitudinal line adjacent to the bottom was defined to monitor the position of the wavefront. Output data were captured at every $25$ time steps. Since this work uses a regularized equation for viscosity (Eq. \ref{eq:bingham_model}), the stoppage of the flow cannot be precisely calculated \citep{Frigaard2005}; an augmented Lagrangian method is required for this \citep{Saramito2017}. Therefore, the final velocity of the runout was computed by using the runout curve and a criteria of $1$ mm/s was attributed to the stoppage, i.e. if the final velocity is below $1$ mm/s, the scenario reached a \textit{quasi-steady} state. However, due to computational cost, some scenarios did not reach the quasi-steady state and only transient behavior was evaluated for these scenarios.

\begin{figure}[H]
    \centering
    \includegraphics[width=0.9\linewidth]{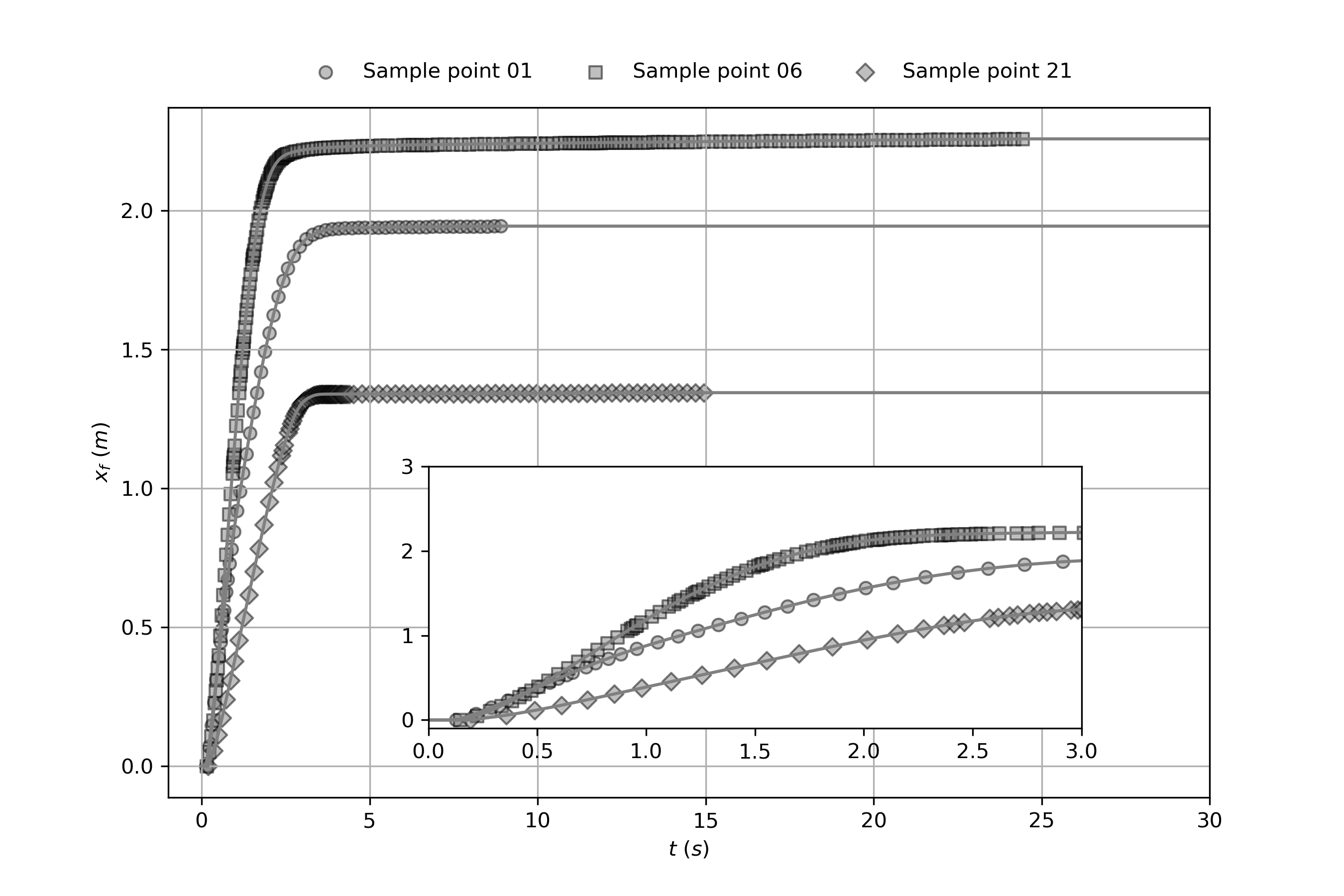}
    \caption{Linear interpolation of runout data $\runout$ over time $t$.}
    \label{fig:interpolation}
\end{figure}

In total, 226 scenarios were explored, where each scenario contains discrete data of runout position over the correspondent time steps. Since a variable time-stepping scheme in function of CFL is applied, time step distribution differs from scenario to scenario. In order to standardize the time domain for every explored scenario, a linear interpolation was employed and $5000$ points were used to represent the interpolated runout data. Figure \ref{fig:interpolation} shows three example scenarios to demonstrate the employed linear interpolation. $183$ scenarios were considered valid to the quasi-steady state analysis, based on the stoppage criteria discussed in Subsection \ref{subsec:numerical}.

\subsection{Uncertainty quantification}
\label{subsec:UQ}

Here, consistent probability density functions (PDFs) are specified to the random input variables of the model, containing the previously known information about them. This step is described in item \ref{subsubsec:probabilistic_modeling}. Then, dimensionality reduction is approached in item \ref{subsubsec:PCA}, which uses a linear basis expansion, generating a set of few basis coefficients and basis components. Item \ref{subsubsec:GPR} describes the metamodeling technique chosen for this work, the Gaussian process regression (GPR) technique, where the few basis coefficients and correspondent input set are used to train a metamodel (or surrogate model); therefore, fast and accurate predictions of basis coefficients can be obtained for any input scenario. We refer to \citep{Jia2013, Li2020, Nagel2020, ElGarroussi2022} for similar applications considering high-dimensional outputs. 

\subsubsection{Probabilistic modeling of input variables}
\label{subsubsec:probabilistic_modeling}

By considering the deterministic model defined in subsection \ref{subsec:numerical}, let us consider an input vector of five input variables $\inputVectorRandom = (\yieldstress, \plasticviscosity, \density, \initialheight, \liftingvelocity)$. These variables are treated as random variables and an uniform distribution is attributed to each input variable. The uniform distribution is chosen due to the maximum entropy principle, which states that the most unbiased distribution must be chosen with the given information \citep{Cunha2014}. In this case, this work constraints the input variables to statistical supports showed in Table \ref{Tab:supports}. The upper and lower limits are chosen following the works of \citet{OBrien1988, Major1992, Sosio2007, Blight2009, Sarsby2000, Fitton2007, Dias2017, Machado2017}, which are applicable to a vast range of materials.

\begin{table}[h!]
    \centering
    \begin{tabular}{|c|c|}
        \hline
        Variable & Distribution \\
        \hline
        $\yieldstress$ & $\uniformDist(0.1 , 200)$ \text{Pa} \\
        \hline
        $\plasticviscosity$ & $\uniformDist(0.01 , 15)$ \text{Pa.s} \\
        \hline
        $\density$ & $\uniformDist(1000 , 2650)$ \text{Kg.m\textsuperscript{-3}} \\
        \hline
        $\liftingvelocity$ & $\uniformDist(0.01 , 1)$ \text{m.s\textsuperscript{-1}} \\
        \hline
        $\initialheight$ & $\uniformDist(0.2 , 1)$ \text{m} \\
        \hline
    \end{tabular}
    \caption{Input variables and their distributions.}
    \label{Tab:supports}
\end{table}

Following the attributed probability distributions, one can construct a design of experiments (DoE) $\DoE = \DoEset$ with $\DoEpoint{\indPoints} \in \domain, \ \domain \subseteq \R^{\nvar}$, $\indPoints=1, \dots, \nDoE$, where $\nDoE$ is the size of DoE and $\nvar$ is the number of input dimensions ($\nvar = 5$). Therefore, each $\DoEpoint{\cdot}$ represents a scenario with $\nvar$ values, corresponding to each input variable. A Latin-Hypercube sampling technique was employed to sample the DoE with size $\nDoE = 226$, limited by the computational cost of each deterministic simulation. 

The simulations were performed according to the deterministic framework (Sec. \ref{subsec:numerical}) and the quantity of interest evaluated was the interpolated wavefront position (runout) over time, or $\runout(t)$. In this case, the number of output dimensions is $\nOutputDimensions=5\,000$, i.e. the number of points used for the interpolation of each curve. The high-dimensionality of the runout-curve problem requires a dimensionality-reduction technique to proceed with the UQ approach. The next item details the dimensionality-reduction strategy.

\subsubsection{Dimensionality reduction}
\label{subsubsec:PCA}

Principal component analysis (PCA) is a data-driven dimensionality-reduction technique, which represents linearly high-dimensional data within a reduced space. For more details, we refer to \cite{Abdi2010}. Let $\dataMatrixPCA = \left[ \outputPoint^{(1)}, \dots, \outputPoint^{(\nDoE)} \right]^{\top} \in \R^{\nDoE \times \nOutputDimensions}$, where $\outputPoint^{(\indPoints)} = \outputPoint(\inputPoint^{(\indPoints)})$ is the $\indPoints$-th high-dimensional output obtained by the deterministic simulator. The dimension-wise mean $\dimensionMean \in \R^{\nOutputDimensions}$ is calculated as:
\begin{equation}
\label{Eq:dimension_wise_mean_chapter3}
    \dimensionMean = \frac{1}{\nDoE} \sum^{\nDoE}_{\indPoints=1} \outputPoint^{(\indPoints)}
\end{equation}

Centered data $\centeredData \in \R^{\nDoE \times \nOutputDimensions}$ is obtained by subtracting $\overline{\outputPoint}$ from each row of $\dataMatrixPCA$. Then, the empirical covariance matrix $\covarianceMatrixPCA \in \R^{\nOutputDimensions \times \nOutputDimensions}$ is calculated based on $\centeredData$:
\begin{equation}
\label{Eq:covariance_matrix_PCA_chapter3}
    \covarianceMatrixPCA = \frac{1}{\nDoE - 1} \centeredData^{\top} \centeredData
\end{equation}

The eigenvalue problem $\covarianceMatrixPCA \basis_{\indBasis} = \lambda_{\indBasis} \basis_{\indBasis}$ is solved and the eigenvalues are sorted in decreasing order $\lambda_1 \geq ... \geq \lambda_{\nOutputDimensions}$, representing the data variance explained by the correspondent eigenvectors. The eigenvector associated with the $\indBasis$-th eigenvalue is denoted by $\basis_{\indBasis} = (\basis_{\indBasis,1}, \dots, \basis_{\indBasis, \nOutputDimensions})^{\top} \in \R^{\nOutputDimensions}$. The truncated PCA basis is obtained by retaining the first $\nbasis$ eigenvectors, with $\nbasis \ll \nOutputDimensions$, and is denoted by $\basisMatrixPCA = (\basis_{\indBasis, \indOutputs})_{1 \leq \indBasis \leq \nbasis, \ 1 \leq \indOutputs \leq \nOutputDimensions} \in \R^{\nbasis \times \nOutputDimensions}$, where the rows of $\basisMatrixPCA$ contain the retained basis vectors. The PCA basis coefficients are obtained by projecting the centered data onto the truncated basis $\coefficientsMatrixPCA = \centeredData \basisMatrixPCA^{\top} \in \R^{\nDoE \times \nbasis}$. Equivalently, $\coefficientsMatrixPCA = ( \coef_{\indBasis}(\inputPoint^{(\indPoints)}) )_{1 \leq \indPoints \leq \nDoE, \ 1 \leq \indBasis \leq \nbasis}$, where the $\indPoints$-th row of $\coefficientsMatrixPCA$ contains the basis coefficients associated with the
input scenario $\inputPoint^{(\indPoints)}$: $\coef(\inputPoint^{(\indPoints)}) = (\coef_1(\inputPoint^{(\indPoints)}),\ldots,\coef_{\nbasis}(\inputPoint^{(\indPoints)}))^\top
\in \R^{\nbasis}$. Finally, the $\indOutputs$-th output dimension of the $\indPoints$-th original series can be represented as:
\begin{equation}
\label{Eq:PCA_reconstruction_chapter3}
    \outputPoint_{\indOutputs}^{(i)} \approx \dimensionMean_{\indOutputs} + \sum^{\nbasis}_{\indBasis=1} \coef_{\indBasis} (\inputPoint^{(i)}) \basis_{\indBasis,\indOutputs}
\end{equation}

All output dimensions for the $\indPoints$-th map can be computed directly in vector form as follows: 
\begin{equation}
\label{Eq:PCA_reconstruction_map_chapter3}
    \outputPoint^{(\indPoints)} \approx \dimensionMean + \basisMatrixPCA^{\top} \coef(\inputPoint^{(\indPoints)})
\end{equation}

In this reduced-order framework, high-dimensional time series can be fast and accurately predicted by metamodeling basis coefficients. The scalar GPR metamodeling technique is described below.

\subsubsection{Metamodeling by Gaussian process regression}
\label{subsubsec:GPR}

Gaussian process regression (GPR) is a probabilistic metamodeling technique used to approximate the relationship between an input vector and a scalar model response. For more details, we refer to \cite{Williams1995}. Let $\deterministicModelGPR : \ \mathbb{X} \subseteq \R^\nvar \longrightarrow \R$ be a multivariate function, modeling a deterministic computer code. As a prior knowledge, assume that $\deterministicModelGPR$ is one trajectory of a Gaussian process $Z$ on $\mathbb{X}$, denoted as $Z \sim \mathcal{GP}(\meanGPR,\kernelGPR)$, where $\meanGPR$ is the mean function and $\kernelGPR$ is the covariance function, or kernel. This prior information can be updated using a DoE $\DoE$ and respective observations $y^{obs}_{\indPoints} = \deterministicModelGPR(\inputPoint^{(\indPoints)})$. This result in a posterior random process, that is proved to be still Gaussian. That conditional GP is denoted $\condGP \sim \mathcal{GP}(\meanGPR_c, \kernelGPR_c)$, where $\meanGPR_c$ and $\kernelGPR_c$ are given in closed-form by:
\begin{equation}
\begin{aligned}
\ConditionalMean(\inputPoint) &= \meanGPR(\inputPoint) + \KernelVectorRow{\inputPoint} \KernelMatrix^{-1} {(y^{obs} - \meanGPR(\DoE))}  \qquad (\inputPoint \in \mathbb{X}) \\
\ConditionalKernel(\inputPoint, \inputPoint') &= \KernelFunction(\inputPoint, \inputPoint') - \KernelVectorRow{\inputPoint} \KernelFunction(\DoE, \DoE)^{-1} \KernelVector{\inputPoint'} \qquad (\inputPoint, \inputPoint' \in \mathbb{X})
\end{aligned}
\label{eq:kriging_mean_variance_chapter3}
\end{equation}
where {$y^{obs} \in \R^\nvar$ is the vector of observed output values}, $\KernelMatrix = (\KernelFunction(\inputPoint^{(\indPoints)}, \inputPoint^{(j)}))_{1 \leq \indPoints,j \leq \nDoE}$ is the covariance matrix at design points, $\KernelVector{\inputPoint} = (\KernelFunction(\inputPoint^{(\indPoints)}, \inputPoint))_{1 \leq \indPoints \leq \nDoE}$ is the column vector of covariances between a new point $x$ and the design points, and $\KernelVectorRow{\inputPoint} = \KernelVector{\inputPoint}^\top$.

In the particular case of high-dimensional output prediction, the scalar response modeled by GPR corresponds to a basis coefficient, i.e $\deterministicModelGPR_\indBasis(\inputPoint) = \coef_{\indBasis}(\inputPoint)$ for each $\indBasis$-th retained basis component. The associated observation vector is the $\indBasis$-th column of the coefficient matrix: $\coefficientsMatrixPCA_{.,\indBasis} = (\coef_{\indBasis}(\inputPoint^{(\indPoints)}))_{\indPoints=1,\dots,\nDoE} \in \R^{\nDoE}$. One scalar GPR metamodel is then trained for each coefficient function $\inputPoint \mapsto \coef_\indBasis(\inputPoint)$, using the DoE $\DoE$ and the observations $\coefficientsMatrixPCA_{\cdot,\indBasis}$ as training data. Thus, a set of $\nbasis$ independent GPR metamodels is constructed: $Z_n^{(\indBasis)} \sim \mathcal{GP}(\ConditionalMean^{(\indBasis)}, \ConditionalKernel^{(\indBasis)})$, with $\indBasis=1,\dots, \nbasis$. 

In this work, we use the Matérn 5/2 kernel and the hyperparameters are estimated through maximum likelihood estimation. The functions and procedures for GPR were done by using the Python library ``gpflow'' \citep{Matthews2017}. For validation, Nash-Sutcliffe efficiency (or $Q^2$ metric) is applied to every output dimension of the original output space \citep{Marrel2011, Sao2026}. Finally, the GPR metamodeling allows the application of global sensitivity analysis, which usually requires a considerable amount of data. The next item presents the basic concepts and which methods this work employs. 

\subsubsection{Algorithm for fast prediction of time series using PCA and GPR}

In the strategy employed by this work, PCA provides a linear reduced space defined by the most important directions of data variability, i.e. the PCA basis components. High-dimensional data is represented in the reduced space by PCA basis coefficients; this representation allows $\nOutputDimensions$-sized high-dimensional data to be described by using $\nbasis$ components, where $\nbasis \ll \nOutputDimensions$. While the PCA basis components are deterministic, the PCA coefficients depend on the input points $\inputPoint$. This fact allows to construct fast-to-evaluate GPR metamodels to map unexplored input points to the correspondent predicted PCA basis coefficients. Finally, the predicted PCA basis coefficients can be decoded by using the PCA basis components and provide an approximate reconstruction of the predicted high-dimensional time series in the original domain.
\begin{algorithm}[H]
\caption{Fast prediction of a new high-dimensional map using PCA and GPR.}
\label{algo:prediction_chapter3}
\begin{enumerate}[label=\arabic*:,leftmargin=2em,nosep]

\item \textbf{Input:} DoE $\DoE$; training data $\dataMatrixPCA$; number of basis components $\nbasis$; input scenario $\inputPoint^{*}$.

\item Compute the dimension-wise mean $\dimensionMean$ with Eq. \ref{Eq:dimension_wise_mean_chapter3} and center training data, obtaining $\centeredData$.

\item Calculate the empirical covariance matrix $\covarianceMatrixPCA$ with Eq. \ref{Eq:covariance_matrix_PCA_chapter3}.

\item Solve the eigenvalue problem $\covarianceMatrixPCA \basis_{\indBasis} = \lambda_{\indBasis} \basis_{\indBasis}$.

\item Sort the eigenvalues $\lambda_1 \geq \dots \geq \lambda_{\nOutputDimensions}$ and associated PCA basis components.

\item Retain the first $\nbasis$ PCA basis components and define the truncated PCA basis $\basisMatrixPCA$.

\item Project $\centeredData$ onto the PCA basis: $\coefficientsMatrixPCA = \centeredData \basisMatrixPCA^{\top}$.

\item \textbf{for} $\indBasis = 1$ \textbf{to} $\nbasis$ \textbf{do}
  \item \hspace*{1em}Define $\coefficientsMatrixPCA_{.,\indBasis} = (\coef_{\indBasis}(\inputPoint^{(\indPoints)}))_{\indPoints=1,\dots,\nDoE}$.
  \item \hspace*{1em}Train the $\indBasis$-th GPR metamodel $\GaussianProcess_{n}^{(\indBasis)}$ using $\DoE$ and $\coefficientsMatrixPCA_{.,\indBasis}$.
  \item \hspace*{1em}Predict $\widehat{\coef}_{\indBasis}(\inputPoint^{*})$ using $\GaussianProcess_{n}^{(\indBasis)}$.
\item \textbf{end for}

\item Define the predicted basis coefficient vector $\widehat{\coef}(\inputPoint^*) = (\widehat{\coef}_1(\inputPoint^*), \dots, \widehat{\coef}_{\nbasis}(\inputPoint^*))^{\top}$.

\item Reconstruct $\widehat{\outputPoint}(\inputPoint^{*}) = \dimensionMean + \basisMatrixPCA^{\top} \widehat{\coef}(\inputPoint^*)$.

\item \textbf{return} $\widehat{\outputPoint}(\inputPoint^{*})$

\end{enumerate}
\end{algorithm}

\subsection{Global sensitivity analysis}
\label{subsec:GSA_chap3}

Global sensitivity analysis (GSA) is useful to identify the influence of input uncertainty over outputs through sensitivity indices. In this work, a variance-based strategy is employed, where uncertainty is reflected by the variance of data. A brief background about GSA is given in this section. We refer to \citep{DaVeiga2021} and \cite{Saltelli2008} for more information about GSA.

\subsubsection{Variance decomposition of data}

Let $\inputVectorRandom = (\inputPointRandom{1}, \dots, \inputPointRandom{\nvar}) \in \R^{\nvar}$ be the vector of independent random input variables of size $\nvar$ and $f(\inputVectorRandom)$ be the model response. Each input variable has uncertainties associated to it and we model them as random variables, denoting $\probDist_{\inputVectorRandom}$ as the probability distribution of $\inputVectorRandom$. Based on the Sobol-Hoeffding decomposition and under suitable conditions, the total variance of $f(\inputVectorRandom)$ can be decomposed as a hierarchical sum of terms associated with individual inputs and their interactions. The variance decomposition formula is written as:
\begin{equation}
\label{eq:variance_decomposition_chapter3}
    \Var(\deterministicModel(\inputVectorRandom)) = \sum_{i=1}^{\nvar} \SobolUnnormalized{i} + \sum_{1\leq i< j \leq \nvar} \SobolUnnormalized{i,j} + \dots + 
    \SobolUnnormalized{1,\dots,\nvar}
\end{equation}

where $\SobolUnnormalized{\indexSet} = \Var(\deterministicModel_{\indexSet}(\inputVectorRandom_{\indexSet}))$ is the partial variance associated to the index set of input variables $\indexSet \subseteq \dimInt$.

The influence of the group of input variables $\inputVectorRandom_{\indexSet}$ can be quantified by the variance explained by $f_{\indexSet}(\inputVectorRandom_{\indexSet})$. We call this quantity \textit{Sobol' index}: if normalized by the total variance, we represent it as $\SobolNormalized{(\cdot)}$; otherwise, we represent the unnormalized index (also called partial variance) as $\SobolUnnormalized{(\cdot)}$. The overall variance is represented by $\totalVariance$, without any subscript. Next, expressions of Sobol' indices and the different types of indices (first-order, closed, total and second-order) are presented along with a brief interpretation for two input variables.

\subsubsection{Interpretation of (unnormalized) Sobol' indices}

Let $\indexSet \subseteq \{ 1, \dots, \nvar \}$ be a non-empty index set and subscript $(\cdot)_{\exceptIndexSet}$ to indicate that variables from the subset $\indexSet$ were excluded from the whole set. The following equations are associated to the group variables $\inputVectorRandom_\indexSet$:

\begin{itemize}
    \item First-order Sobol' index: $\SobolUnnormalized{i} = \Var \left( f_{i}(\inputVector_{i}) \right)$, where $i \in \{1,\dots,\nvar\}$. It represents the isolated contribution (or main effect) of $\inputVectorRandom_i$ over output variability.
    
    \item Closed Sobol' index: $\sobolClosedD{\indexSet} = \Var \left( \Esp \left[ f(\inputVectorRandom) \vert \inputVectorRandom_{\indexSet} \right] \right) $. If $\indexSet = \{ i,j \}$, it contains the isolated effects of $\inputVectorRandom_i$ and $\inputVectorRandom_j$ and the interaction between both variables; therefore, $\sobolClosedD{i,j} = \SobolUnnormalized{i} + \SobolUnnormalized{j} + \SobolUnnormalized{ij}$. For $\indexSet = \{ i \}$, $\sobolClosedD{i} = \SobolUnnormalized{i}$.

    \item  Total Sobol' index: $\sobolTotD{\indexSet} = \totalVariance - \sobolClosedD{\exceptIndexSet}$. If $\indexSet = \{ i,j \}$, it represents the main effect of both variables and the interaction of $\{ \inputVectorRandom_i, \inputVectorRandom_j\}$ with the remaining ones $\inputVectorRandom_{\exceptIndexSet}$.
\end{itemize}

\subsubsection{Estimation of sensitivity indices in the context of high-dimensional outputs}

The estimation of sensitivity indices in a high-dimensional context is detailed in \citep{Sao2026}, we provide herein a brief background. Among many possible techniques, \textit{pick-freeze} (PF) schemes or estimators are used herein due to their nice statistical properties and for being general. In a PF scheme, one considers two independent random vectors $X$ and $Z$ drawn from $\probDist_X$. The model output is evaluated twice: firstly, by computing $Y= f(X)$; and secondly, by fixing (freezing) the coordinates of $X$ in $I$ and choosing (picking) the other coordinates (denoted ${(\cdot)}_{\exceptIndexSet}$) in $Z$, leading to $Y^{*} = f(\inputVectorRandom_{\indexSet}, Z_{\exceptIndexSet})$. By definition of $X$ and $Z$, outputs $Y$ and $Y^*$ have the same probability distribution. But they are now dependent and their correlation is proved to be equal to the closed Sobol' index. Furthermore, by using $\Cov(Y, Y^*) = \Esp[Y Y^*] - f_0^2$ and $\Var(Y) = \Esp[Y^2] - f_0^2$, with $f_0 = \Esp \left[ \frac{Y + Y^*}{2} \right]$, the following equation can be written:

\begin{equation}
    \sobolClosed{\indexSet} = \frac{\Cov(Y, Y^{*})}{\Var ( Y )} = \frac{ \Esp[ Y Y^{*} ] - \left( \Esp \left[ \frac{Y + Y^*}{2} \right] \right)^2 }{ \Esp \left[ \frac{Y^2 + (Y^*)^2}{2} \right] - \left( \Esp \left[ \frac{Y + Y^*}{2} \right] \right)^2}
\label{Eq.closed_Sobol_variance_form_chapter3} 
\end{equation}

Based on Eq. \ref{Eq.closed_Sobol_variance_form_chapter3}, the Sobol' indices can be estimated empirically. A convenient choice is the Janon-Monod estimator, which is consistent and asymptotically efficient \citep{Janon2014a, Janon2014b}. In a similar way, total Sobol' indices can be computed through the Jansen's estimator \citep{Jansen1999}. For vector-valued PF estimation, the same logic is applied. However, the outputs $Y$ and $Y^{*}$ are vector-valued in this case, so matrix multiplications must be employed instead of simple scalar multiplications. The vector-valued PF estimation is particularly useful to obtain \textit{sensitivity curves} or \textit{sensitivity maps}, which represent the Sobol' indices associated to each individual output, denoted as $\sobolClosed{\indexSet}(\outputHighDimensional)$. As demonstrated in \citet{Sao2026}, sensitivity curves can be computed as:
\begin{equation} 
\label{eq:closedIndexWithCoef_chapter3}
\pfEstim{\sobolClosed{\indexSet}}(\outputHighDimensional(\inputVectorRandom)) = \frac{\basisVec^\top \, \pfEstim{\sobolClosedD{\indexSet}}(\coef(\inputVectorRandom)) \, \basisVec}{\basisVec^\top \, \pfEstim{\Cov}(\coef(\inputVectorRandom)) \basisVec}
\end{equation}

where $\pfEstim{\sobolClosedD{\indexSet}}(\coef(\inputVectorRandom)) \in \R^{\nbasis \times \nbasis}$ and $\pfEstim{\Cov}(\coef(\inputVectorRandom)) \in \R^{\nbasis \times \nbasis}$ are vector-valued PF estimation of covariance matrices associated with the vector of basis coefficients.

Finally, concise sensitivity indices can be defined, such as the generalized sensitivity indices (GSIs) \citep{Lamboni2009, Lamboni2011, Perrin2021}. They represent variance-weighted averages of the Sobol’ indices, providing one single scalar value to quantify the influence of a given variable or their interaction: 
\begin{equation}
    \label{eq:gsi_estimator_chapter3}
    \GSI_\indexSet = \frac{\Tr{\pfEstim{\sobolClosedD{\indexSet}} (\coef(\inputVectorRandom)}}{\Tr{\pfEstim{\Cov}(\coef(\inputVectorRandom))}}
\end{equation}

where $\Tr{\cdot}$ denotes the trace of a matrix.

\section{Results and discussion}
\label{sec:results_chap3}

First, we present the analysis of deterministic data in Subsection \ref{subsec:deterministic_results} with a preliminary statistical approach. Then, GSA analysis and further physical interpretation are presented in Subsection \ref{subsec:sensitivity_analysis_results}.

\subsection{Analysis of the deterministic results}
\label{subsec:deterministic_results}

We start by evaluating quasi-steady runout data. For this analysis, we define the following two non-dimensionalized numbers. First, the Bingham number $\binghamnumber$, which represents a ratio between the yield stress and initial hydrostatic pressure. Second, the inertial number $\inertianumber$, relating a typical dynamic pressure caused by inertia with the yield stress. Both are defined in Eq. \ref{eq:non_dimensionalized_numbers}.  

\begin{equation}
    \label{eq:non_dimensionalized_numbers}
    \binghamnumber = \frac{\yieldstress}{\density g \initialheight} \ ; \ \inertianumber = \frac{\density V_{med}^2}{\yieldstress}
\end{equation}

where $V_{med}$ is a mean velocity, defined by the stoppage time $T_{stop}$ and the quasi-steady runout $\runout$.

\begin{figure}[H]
    \centering
    \begin{subfigure}[b]{0.48\textwidth}
        \centering
        \includegraphics[width=\textwidth]{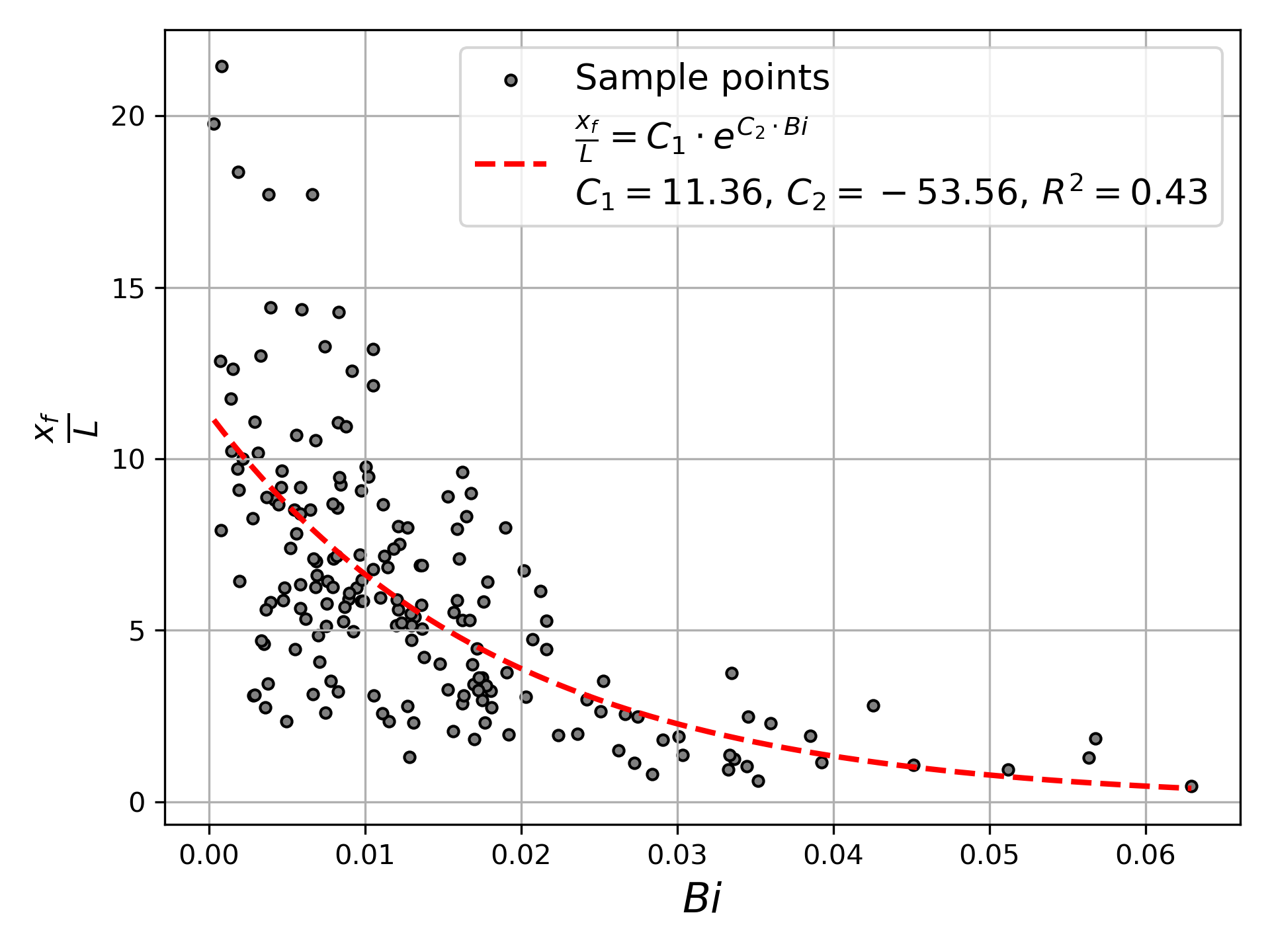}
        \caption{Non-dimensionalized steady-state runout $\frac{\runoutsteady}{\lengthreservoir}$ vs. Bingham number $\binghamnumber$. Dashed red line represents an exponential trend line.}
        \label{fig:runout_bingham}
    \end{subfigure}
    \hfill
    \begin{subfigure}[b]{0.48\textwidth}
        \centering
        \includegraphics[width=\textwidth]{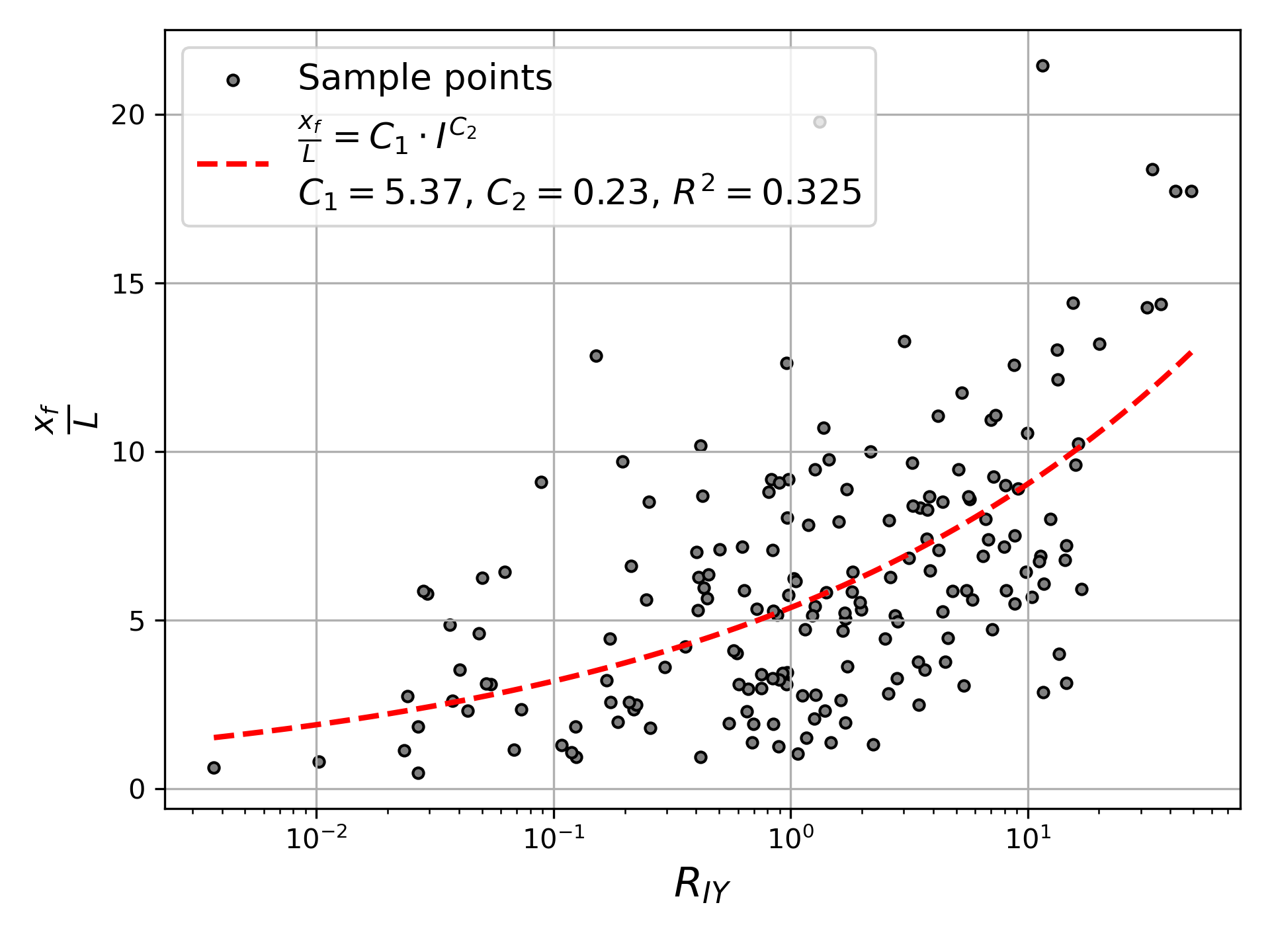}
        \caption{Non-dimensionalized steady-state runout $\frac{\runoutsteady}{\lengthreservoir}$ vs. Inertia number $\inertianumber$. Dashed red line represents a power-law trend line.}
        \label{fig:runout_inertia}
    \end{subfigure}
    \caption{Comparison of runout behavior with respect to Bingham number and inertia number.}
    \label{fig:runout_comparison}
\end{figure}

Figure \ref{fig:runout_bingham} shows the steady-state runout points $\runoutsteady$ normalized by the length of the reservoir $\lengthreservoir$ of each correspondent scenario in function of the Bingham number $\binghamnumber$. An exponential trend line represented by the dashed red line shows an inverse relationship, i.e. as $\binghamnumber$ increases, the runout decreases. It indicates that $\yieldstress$ reduces the effect of potential energy stored in initial hydrostratic pressure, resulting in shorter runouts (e.g. $\binghamnumber > 0.025$). For smaller values of $\binghamnumber$, the runout tends to be longer, since the flow has more potential energy in comparison with $\yieldstress$. However, the significant variability of points for $\binghamnumber < 0.025$ indicates that other effects play significant role over the runout, such as the inertial effects.

Figure \ref{fig:runout_inertia} presents the runout behavior in function of the inertial number $\inertianumber$. The power-law trend indicates a direct relationship between runout and inertia, i.e. the runout increases as the inertia increases. Inertia increases if $\plasticviscosity$ and $\yieldstress$ are small, since less resistive viscous forces are counter-acting the inertia of the flow. On the other hand, higher values of $\initialheight$ and $\liftingvelocity$ contribute to inertia, since more potential energy is initially available due to $\initialheight$, which can be converted into kinetic energy in function of the lifting velocity of the gate.

\begin{figure}[H]
    \centering
    \includegraphics[width=0.95\linewidth]{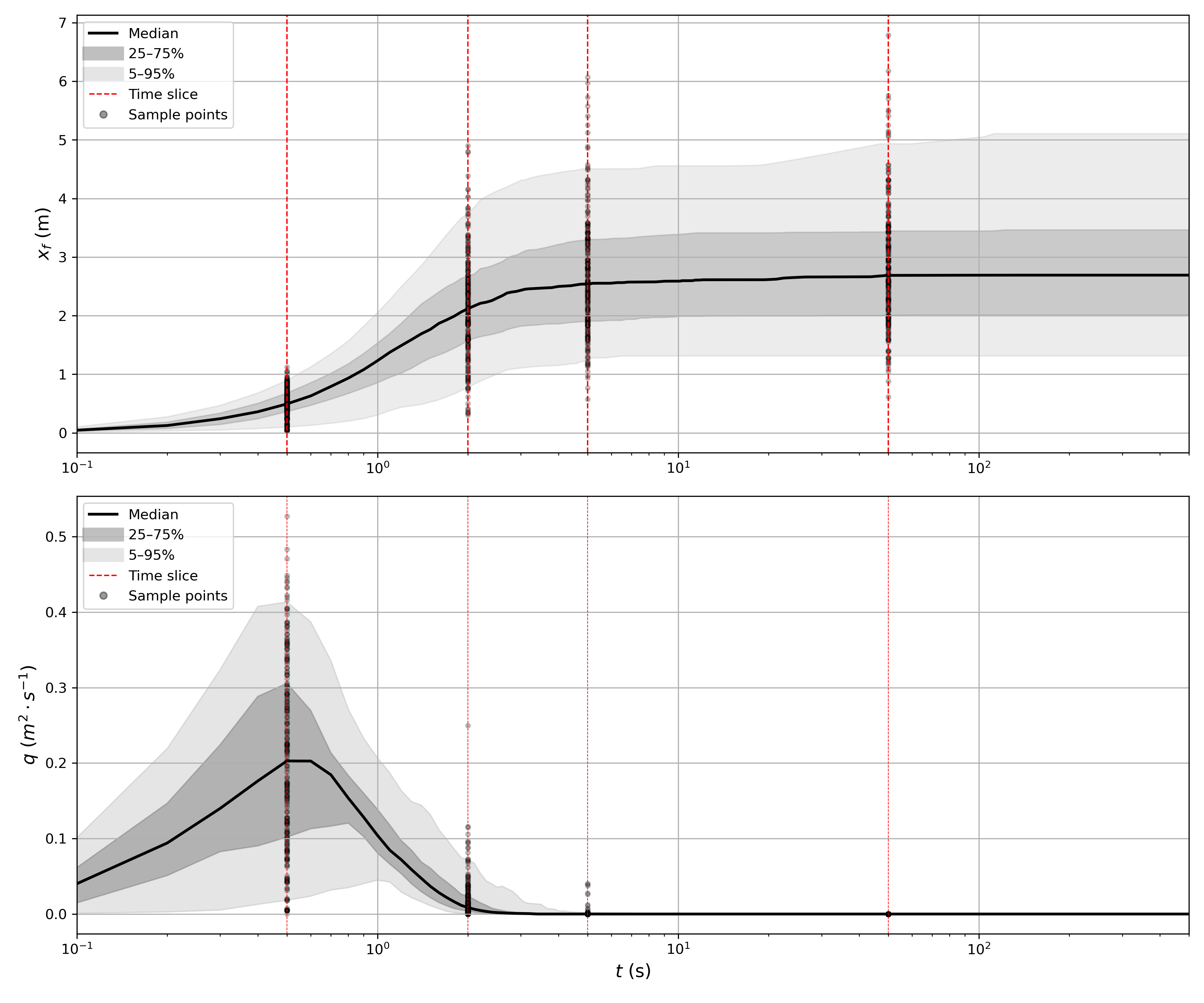}
    \caption{Distribution of $226$ data points of runout $\runout(t)$ (top) and flow rate $\flowrate$ (bottom) over time. Full curves represent the median curves and the shaded areas correspond to the $25-75 \%$ (darker) and $5-95\%$ (lighter) quantile envelopes.}
    \label{fig:distribution_curves_runout}
\end{figure}

Both Figs. \ref{fig:runout_bingham} and \ref{fig:runout_inertia} show the steady-state results in function of $\binghamnumber$ and $\inertianumber$. However, it is clear that the quasi-steady runout is not exclusively a function of $\binghamnumber$ nor $\inertianumber$ due to the low $R^2$ of both regression functions. To further investigate the case, we may explore the transient behavior of the runout and hydrographs. Figure \ref{fig:distribution_curves_runout} shows the development of runout (top) and flow rate (bottom) over time. The results are shown in terms of median and envelopes that encompass 50\% and 90\% of data. By observing Fig. \ref{fig:distribution_curves_runout}, we note that the median runout curve follows the typical shape of a rapid development, which is coherent with the peak of median hydrograph, and then a tendency to an arrested state.

\subsection{Global sensitivity analysis}
\label{subsec:sensitivity_analysis_results}

\begin{figure}[H]
    \centering
    \includegraphics[width=0.95\linewidth]{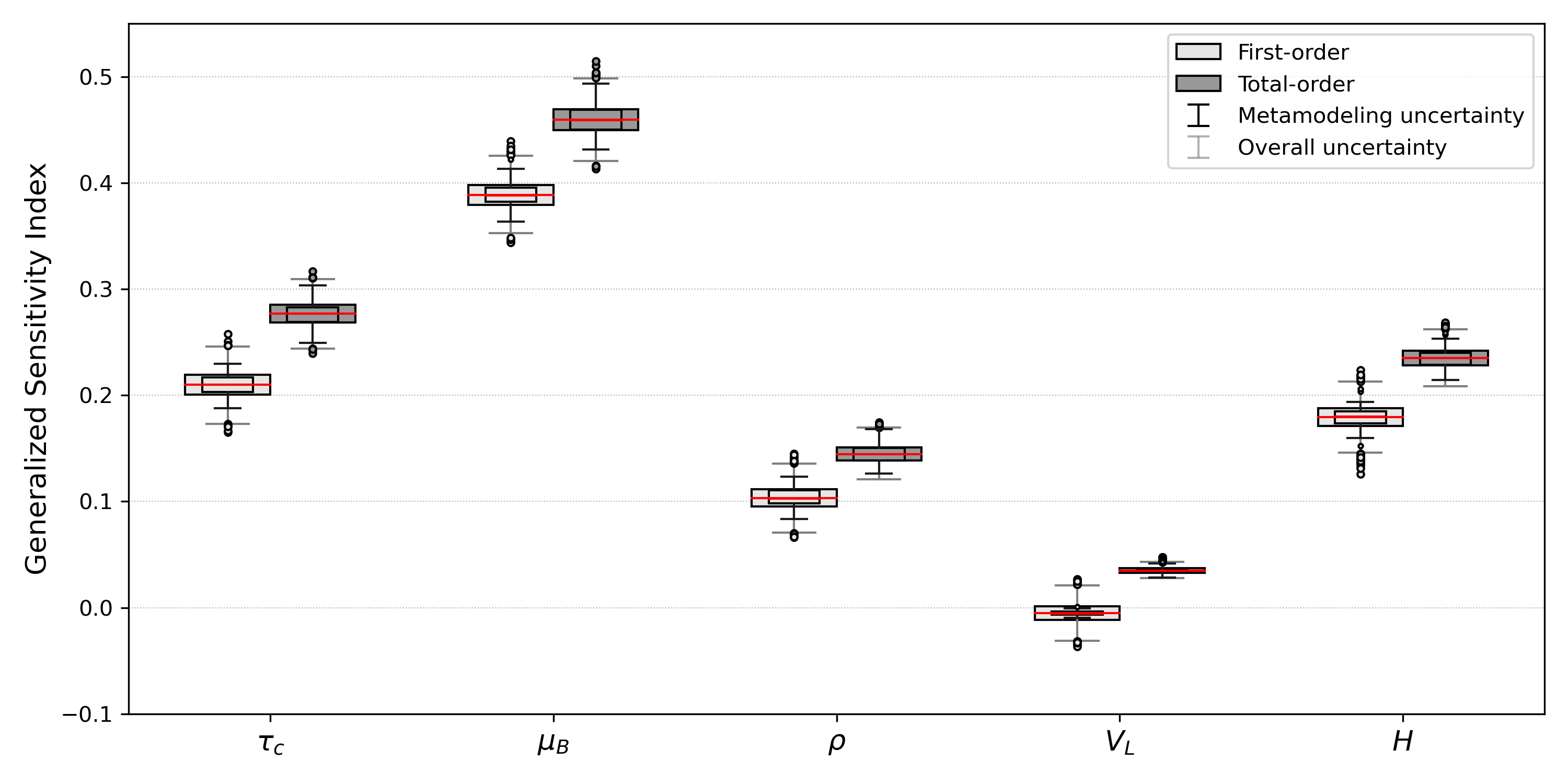}
    \caption{First- and total-order generalized sensitivity indices (GSI) of each input parameter. Red lines refer to the median (50th percentile) and the box spans the interquartile range (IQR), where the lower edge represents the first (25th percentile) quartile and the upper edge represents the third (75th percentile) quartile. The whiskers represent the range of data within $1.5 \times \text{IQR}$.}
    \label{fig:GSI}
\end{figure}

    Global sensitivity analysis was performed by considering the input parameter space and the resulting 226 runout curves. We may start by showing the GSIs in Fig. \ref{fig:GSI}, whose calculation and meaning are given by Eq. \ref{eq:gsi_estimator_chapter3}. Each input variable is associated with first- and total-order GSIs, whose values are synthesized into a boxplot representation. It is observable that $\plasticviscosity$ is the most important input variable, followed sequentially by $\yieldstress$, $\initialheight$, $\density$ and $\liftingvelocity$. We can also notice that the median values of first-order indices are close to the values of total-order indices, presenting even overlay of the whiskers that represent the uncertainty in the calculation. Therefore, interaction effects between variables are weak in this problem. 

\begin{figure}[H]
    \centering
    \includegraphics[width=0.8\linewidth]{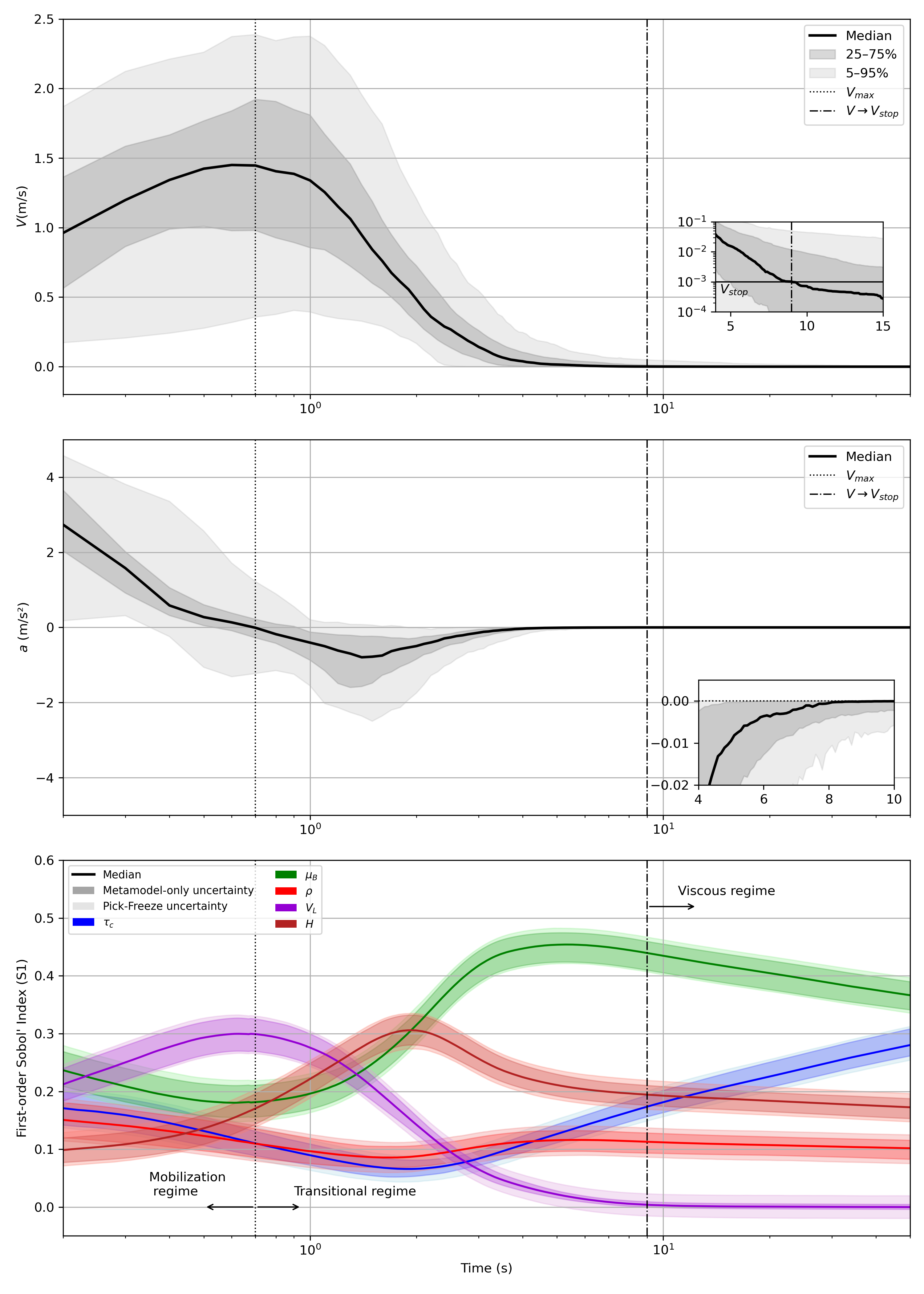}
    \caption{Time evolution of wavefront velocity (top), acceleration (middle), and first-order Sobol' indices for the runout position (bottom). The top and middle panels show the median (black line), interquartile (25–75\%) and 5–95\% envelopes in gray. Vertical lines indicate the time of maximum velocity ($V_\mathrm{max}$, dotted) and the time when the median velocity reaches the threshold $V_\mathrm{stop}$ (dot-dashed). Shaded areas represent uncertainties due to the metamodel (light shade) and the pick-freeze estimation (dark shade). Three flow regimes are annotated: mobilization, transitional, and viscous.}
    \label{fig:Sobol}
\end{figure}

In order to investigate with more details the sensitivity of input parameters over runout, we present the temporal analyses of the flow in Fig. \ref{fig:Sobol}. Three analyses are shown: the wavefront velocity $V(t)$, the wavefront acceleration $a(t)$ and the first-order Sobol' indices defined in the time domain (Eq. \ref{eq:closedIndexWithCoef_chapter3}). We opted to show the first-order effect due to the lack of significative interactions, which are denoted by $S_{(\cdot)}$, where $(\cdot)$ may refer to the input parameter of interest. By observing the top and middle panels, it is noticed that the flow accelerates until it reaches its peak velocity at $t \approx 0.7 \ s$ and then decelerate tending to an arrested state at $t \approx 9 \ s$. Based on these time instants, we can divide the flow roughly into 3 phases: an inertial phase, a transition phase and a viscous phase. We also recall that the driving force parameters are $\liftingvelocity$, $\density$ and $\initialheight$ and the resistive parameters are $\yieldstress$ and $\plasticviscosity$, due to the direct and inverse correlations with runout, respectively. Contrary to the GSIs, this representation allows us to evaluate the sensitivity of each input parameter at every time instant. Now, each phase will be separately discussed and physically interpreted.  

\subsubsection{Inertial (mobilization) phase}

The inertial phase occurs at $t \approx [0.0, 0.7] \ s$. By observing $a(t)$ in the middle panel of Fig. \ref{fig:Sobol}, we note that this is an acceleration phase, where the driving forces are able to surpass the viscous resistance. Therefore, \textit{inertial} effects dominate at this time range. In the bottom panel, $\sobollifting$ is the most significant input parameter and it reaches its peak at $t \approx 0.7 \ s$. This phase is characterized physically by the \textit{mobilization} of stored fluid, which accelerates the flow and is strongly controlled by how fast the gate is lifted; thus, this phase can also be called a \textit{mobilization phase}. Furthermore, this mobilization and consequent inertia occur because of the conversion of stored potential energy into kinetic energy. However, this conversion is not instantaneous due to the resistive force of $\plasticviscosity$, which is also important during this phase (see $\sobolviscosity$). 

On the other hand, $\sobolinitial$ is initially low, meaning that $\initialheight$ is not too important for the runout at this phase. Generally, we can recognize $\initialheight$ as a driving-force parameter for two action mechanisms: (i) it defines the quantity of stored potential energy of the fluid; and (ii) it scales with the flow depth, related to the pressure gradient. At the mobilization phase, potential energy is still being converted to kinetic energy, so scenarios with big values of $\initialheight$ are not able to affect the variance of the runout yet, i.e. the fluid was not significantly mobilized. Furthermore, the flow is still developing, so the flow depths are not big enough to exert sufficient pressure gradient to affect the runout. Nevertheless, we can observe that $\sobolinitial$ grows over time within this phase, meaning that the two action mechanisms of $\initialheight$ are increasingly more important.

The other variables $\yieldstress$ and $\density$ also remain unimportant. The typical stresses in the fluid surpass the yield stress, i.e. there is movement, and shear rates tend to be high, which makes the $\yieldstress$ term of apparent viscosity to be low compared to $\plasticviscosity$ (see Eq. \ref{eq:bingham_model}). About $\density$, it scales uniformly with inertia and gravitational terms in the momentum equations, so it does not modify significantly the variance of runout. In short, this phase is mostly dependent of $\liftingvelocity$ and $\plasticviscosity$, with a growing but still unimportant dependence of $\initialheight$.

\subsubsection{Transition phase}

The transition phase is defined at $t \approx [0.7, 9.0] \ s$. Physically, it is characterized by the deceleration of the wavefront (middle panel of Fig. \ref{fig:Sobol}). We can note a sharp decrease of $\sobollifting$, which indicates that inertia effects caused by the mobilization are fading. Meanwhile, $\sobolinitial$ increases due to consequent increase of pressure gradient (bigger flow depths) scaled by $\initialheight$, and conversion of potential energy to kinetic energy still affects the runout. Therefore, $\initialheight$ takes over as the main driving-force parameter, while $\sobollifting$ tends to zero in the later instants of the transition phase. Since the wavefront decelerates, the viscous resistance turns to be predominant, associated with a significative increase of $\sobolviscosity$, reaching its peak at $t \approx 7 \ s$. We can also note an increase of $\sobolyield$, after $t \approx 2 \ s$, although remaining relatively unimportant. After this time, due to the very low velocities, shear rate and shear stress tend to be low. Then, the flow slowly tends to initiate a purely viscous regime.

\subsubsection{Viscous phase}

The viscous phase is defined after $t \approx 9 \ s$, where the median of wavefront velocity reaches $V_{stop}$. The deceleration is almost negligible and a very slow flow takes place. Physically, this phase is governed by two forces, namely the pressure gradient (scaled by $\sim \density g \initialheight$) as driving force and the viscosity as resistance parameter. While $\soboldensity$ and $\sobolinitial$ remains constant, we can notice that a gradual shift of importance between $\sobolviscosity$ and $\sobolyield$, indicating a tendency to the arrested state. For $t \rightarrow \infty$, hydrostatic pressure of the flow final profile is perfectly balanced with $\yieldstress$ \citep{Pashias1996}; therefore, $\yieldstress$ is supposed to be relevant. In fact, empirical methods of yield stress estimation are scaled by the Bingham number (Eq. \ref{eq:non_dimensionalized_numbers}), usually presenting good fits \citep{Coussot1995, Pashias1996, Clayton2003, Saak2004, Nguyen2006, Pereira2022}. 

In this context, the influence of $\plasticviscosity$ may explain the low correlation between $\binghamnumber$ and $\runoutsteady/\lengthreservoir$ presented in Fig. \ref{fig:runout_bingham}. The arrested state was not achieved due to limitation of the regularization technique, described in Subsection \ref{subsec:numerical}. Furthermore, the computational cost to reach an exact arrested state would be too high. Some works report experiments that take up to hours to visibly stop flowing \citep{Coussot1995}, which means that it would require too long simulations. Nevertheless, the explored range of time indicates the tendency of increasing importance of yield stress.

\subsection{Interpretation of results in alternative rheometry context}

The observed results present repercussions in a practical sense of alternative rheometry techniques that use the dam-break flow. As mentioned in Section \ref{sec:dam_break_flows}, these techniques allow for the estimation of rheological parameters based on measurements of runout (or spread) and slump, either during the transient phase or at the final (arrested) state of the flow.

For steady-state tests, also known as stoppage tests, the final spread or slump measurement, once the flow has come to rest, is used to determine the rheological property, especially the yield stress \citep{Pashias1996, Roussel2005a, Coussot2014}. Although the results of this study indicate that viscosity is the most sensitive parameter influencing the spread in the viscous phase, it should be noted that the sensitivity to yield stress shows a trend suggesting growth beyond the time interval analyzed (50 s). Considering these aspects, and given the slow evolution of the spread over time, on the order of one hour until complete arrest of the flow, yield stress is confirmed as the property intrinsically associated with the final slump and spread measurement.

In this sense, the results of this study, as shown in Fig. \ref{fig:Sobol}, indicate that the lifting velocity has no effect on the final spread measurement, supporting the conclusion that models based on slump and spread measurements in steady-state conditions are not influenced by the lifting velocity. This finding is consistent with observations made in works such as Pashias et al. (1996) and Zheng et al. (2024). However, it is important to highlight that the present study analyzed wide parameter ranges, and further analysis focusing on narrower ranges, e.g. those that involve low-viscosity materials, should be conducted to verify whether the lifting velocity, under such conditions, may play a more relevant role in the spread and especially in the slump measurement, which evolves more rapidly and may therefore be more susceptible to lifting effects \citep{Gao2015a, Zhang2024}.

In techniques that use the transient regime for rheological characterization, a commonly employed parameter, whether in the slump test \citep{Ferraris1998,Chidiac2000}, mini-slump \citep{Choi2016}, or channel \citep{Balmforth2007}, is a characteristic time associated with a spatial displacement (slump or spread). The results indicate that the lifting velocity has a significant influence on the development of flow properties in the initial stages, directly affecting the rates of spreading/slump. In other words, the spreading/slump curves can be shifted in time according to the applied velocity.

Considering, for example, the methodology of \citep{Ferraris1998}, which uses the time required for the vertical displacement of 100 mm of the sample to define the plastic viscosity, it is assumed that this rheological parameter is strongly influenced by the lifting dynamics. As shown in the Section \ref{subsec:sensitivity_analysis_results}, this effect directly impacts the transient phase of the flow, potentially altering, depending on the lifting velocity, the time needed for the material to reach the characteristic section of the methodology.

An evaluation by \citet{Pereira2021} indicates that the model of \citet{Ferraris1998} is highly sensitive to the characteristic time of the test, which may change the plastic viscosity by about 6 Pa·s for variations of 0.1 s, considering slump values below 200 mm. Thus, given that slump develops more rapidly than spreading and that the transient phase is strongly influenced by the lifting velocity, inadequate control of this velocity can significantly increase errors in obtaining the rheological parameter.

Similar behavior occurs in methodologies that use the inclined channel. In the Bostwick consistometer \citep{Balmforth2007}, for example, three control sections are employed to define characteristic times directly used in determining rheological properties. The global analysis (Fig. \ref{fig:Sobol}) shows that the lifting velocity significantly influences the development of the flow during the initial times, affecting the measured times at these sections. Therefore, based on the observed results, it can be assumed that higher velocities tend to reduce the initial development time of spreading, underestimating viscosity and yield stress. Consequently, the observations of this study indicate that, in methods based on the transient phase of flow, the lifting velocity is a critical factor that may compromise the estimation of rheological parameters.

\section{Conclusion}
\label{sec:conclusion_chap3}

In this work, we used a global sensitivity analysis (GSA) framework to investigate the case of gradual and idealized dam-break flow of Binghamian fluids, which consists of a reserved fluid being released by the gradual lifting of a gate and flowing downstream a horizontal channel. Five input parameters relevant to the problem were considered: yield stress, plastic viscosity, density, initial height of fluid and lifting velocity of the gate. As quantity of interest, we focused on the evolution of the wavefront position over time, i.e. runout curves. Some conclusions of this work are listed:

\begin{enumerate}

    \item By confronting the physical behavior of the wavefront position with Sobol' indices curves, it is possible to define three regimes. The first regime is an inertial (or mobilization) regime, creating acceleration by the mobilization of fluid controlled by the lifting velocity and balanced by the viscous resistance. The second regime is characterized by the deceleration of the flow, presenting a predominance of the viscous resistance over the driving forces. However, the lifting velocity is still a relevant variable, indicating that mobilization effects still take place along with pressure gradient, represented by the increase of importance of the initial height. Finally, the third regime is a viscous regime, where the viscous resistance completely dominates and balances out the pressure gradient, without interference of lifting velocity.

    \item The GSA suggests that rheological property estimation methods based on the transient phase may be compromised, as they rely on short time intervals that can be directly affected by the mould lifting velocity. This influence may lead to inaccurate parameter estimation, highlighting the need for caution when applying such methods and the importance of isolating the lifting effect during experiments. Conversely, in the steady phase of the flow, yield stress was found to have a substantial influence on the behavior, even when a fully steady state might not have been reached, thus supporting methodologies that rely on the final material configuration (slump or spread) to estimate $\yieldstress$.

     \item The analysis conducted opens new research questions and outlines potential objectives to further advance the understanding of viscoplastic dam-break flows. First, the proposed methodology could be extended to reduced-scale tests, such as the mini-slump, to verify whether the influence of lifting velocity observed in standard slump tests persists at smaller scales and axisymmetric coordinate system. Additionally, conditional GSA may help to identify input subspaces where lifting velocity plays a dominant role, particularly for low-viscosity materials. Finally, another promising direction is to reformulate the analysis in dimensionless terms, allowing result generalization and regime identification across different geometries and materials.

\end{enumerate}

\bibliographystyle{abbrvnat}
\bibliography{references}

@book{Saltelli2008,
author       = {Saltelli, Andrea and Ratto, Marco and Andres, Terry and Campolongo, Francesca and Cariboni, Jessica and Gatelli, Debora and Saisana, Michaela and Tarantola, Stefano},
  title        = {Global sensitivity analysis: The primer},
  year         = {2008},
  publisher    = {John Wiley \& Sons}
}

@book{DaVeiga2021,
author       = {Da Veiga, S{\'e}bastien and Gamboa, Fabrice and Iooss, Bertrand and Prieur, Cl{\'e}mentine},
  title        = {Basics and trends in sensitivity analysis: Theory and practice in R},
  year         = {2021},
  publisher    = {SIAM}
}

@article{Nagel2020,
author       = {Nagel, Joseph B and Rieckermann, J{\"o}rg and Sudret, Bruno},
  title        = {Principal component analysis and sparse polynomial chaos expansions for global sensitivity analysis and model calibration: Application to urban drainage simulation},
  journal      = {Reliability Engineering \& System Safety},
  volume       = {195},
  pages        = {106737},
  year         = {2020},
  publisher    = {Elsevier}
}

@article{Perrin2021,
author       = {Perrin, TVE and Roustant, Olivier and Rohmer, J{\'e}r{\'e}my and Alata, Olivier and Naulin, JP and Idier, D{\'e}borah and Pedreros, Rodrigo and Moncoulon, D and Tinard, P},
  title        = {Functional principal component analysis for global sensitivity analysis of model with spatial output},
  journal      = {Reliability Engineering \& System Safety},
  volume       = {211},
  pages        = {107522},
  year         = {2021},
  publisher    = {Elsevier}
}

@article{Abdi2010,
author       = {Abdi, Herv{\'e} and Williams, Lynne J},
  title        = {Principal component analysis},
  journal      = {Wiley interdisciplinary reviews: computational statistics},
  volume       = {2},
  number       = {4},
  pages        = {433--459},
  year         = {2010},
  publisher    = {Wiley Online Library}
}

@article{Janon2014a,
author       = {Janon, Alexandre and Nodet, Ma{\"e}lle and Prieur, Cl{\'e}mentine},
  title        = {Uncertainties assessment in global sensitivity indices estimation from metamodels},
  journal      = {International Journal for Uncertainty Quantification},
  volume       = {4},
  number       = {1},
  year         = {2014},
  publisher    = {Begel House Inc.}
}

@article{Lamboni2011,
author       = {Lamboni, Matieyendou and Monod, Herv{\'e} and Makowski, David},
  title        = {Multivariate sensitivity analysis to measure global contribution of input factors in dynamic models},
  journal      = {Reliability engineering \& system safety},
  volume       = {96},
  number       = {4},
  pages        = {450--459},
  year         = {2011},
  publisher    = {Elsevier}
}

@article{Jansen1999,
author       = {Jansen, Michiel JW},
  title        = {Analysis of variance designs for model output},
  journal      = {Computer Physics Communications},
  volume       = {117},
  number       = {1-2},
  pages        = {35--43},
  year         = {1999},
  publisher    = {Elsevier}
}

@article{Janon2014b,
author       = {Janon, Alexandre and Klein, Thierry and Lagnoux, Agnes and Nodet, Ma{\"e}lle and Prieur, Cl{\'e}mentine},
  title        = {Asymptotic normality and efficiency of two {S}obol' index estimators},
  journal      = {ESAIM: Probability and Statistics},
  volume       = {18},
  pages        = {342--364},
  year         = {2014},
  publisher    = {EDP Sciences}
}

@article{Li2020,
author       = {Li, Min and Wang, Ruo-Qian and Jia, Gaofeng},
  title        = {Efficient dimension reduction and surrogate-based sensitivity analysis for expensive models with high-dimensional outputs},
  journal      = {Reliability Engineering \& System Safety},
  volume       = {195},
  pages        = {106725},
  year         = {2020},
  publisher    = {Elsevier}
}

@article{Marrel2011,
author       = {Marrel, Amandine and Iooss, Bertrand and Jullien, Michel and Laurent, B{\'e}atrice and Volkova, Elena},
  title        = {Global sensitivity analysis for models with spatially dependent outputs},
  journal      = {Environmetrics},
  volume       = {22},
  number       = {3},
  pages        = {383--397},
  year         = {2011},
  publisher    = {Wiley Online Library}
}

@article{Sao2026,
author       = {Yuri Taglieri S\'ao and Olivier Roustant and Geraldo de Freitas Maciel},
  title        = {Fast pick‐freeze estimation of {S}obol’ sensitivity maps using basis expansions},
  journal      = {Reliability Engineering \& System Safety},
  volume       = {265},
  number       = {Part A},
  pages        = {111504},
  year         = {2026}
}

@article{Williams1995,
author       = {Williams, Christopher and Rasmussen, Carl},
  title        = {{G}aussian processes for regression},
  journal      = {Advances in neural information processing systems},
  volume       = {8},
  year         = {1995}
}

@article{Pashias1996,
author       = {Pashias, N and Boger, DV and Summers, J and Glenister, DJ},
  title        = {A fifty cent rheometer for yield stress measurement},
  journal      = {Journal of rheology},
  volume       = {40},
  number       = {6},
  pages        = {1179--1189},
  year         = {1996},
  publisher    = {The Society of Rheology}
}

@article{Balmforth2007,
author       = {Balmforth, NJ and Craster, RV and Perona, P and Rust, AC and Sassi, R},
  title        = {Viscoplastic dam breaks and the {B}ostwick consistometer},
  journal      = {Journal of non-newtonian fluid mechanics},
  volume       = {142},
  number       = {1-3},
  pages        = {63--78},
  year         = {2007},
  publisher    = {Elsevier}
}

@article{Roussel2005a,
author       = {Roussel, N and Coussot, Philippe},
  title        = {“{F}ifty-cent rheometer” for yield stress measurements: from slump to spreading flow},
  journal      = {Journal of rheology},
  volume       = {49},
  number       = {3},
  pages        = {705--718},
  year         = {2005},
  publisher    = {AIP Publishing}
}

@article{Gao2015a,
author       = {Gao, Jinglong and Fourie, Andy},
  title        = {Spread is better: An investigation of the mini-slump test},
  journal      = {Minerals Engineering},
  volume       = {71},
  pages        = {120--132},
  year         = {2015},
  publisher    = {Elsevier}
}

@article{Pereira2022,
author       = {Pereira, Jo{\~a}o Batista and S{\'a}o, Yuri Taglieri and de Freitas Maciel, Geraldo},
  title        = {Numerical and experimental application of the automated slump test for yield stress evaluation of mineralogical and polymeric materials},
  journal      = {Rheologica Acta},
  volume       = {61},
  number       = {2},
  pages        = {163--182},
  year         = {2022},
  publisher    = {Springer}
}

@article{Ancey2009,
author       = {Ancey, Christophe and Cochard, Steve},
  title        = {The dam-break problem for {H}erschel--{B}ulkley viscoplastic fluids down steep flumes},
  journal      = {Journal of Non-Newtonian Fluid Mechanics},
  volume       = {158},
  number       = {1-3},
  pages        = {18--35},
  year         = {2009},
  publisher    = {Elsevier}
}

@article{Matthews2017,
author       = {Matthews, Alexander G. de G. and
              {van der Wilk}, Mark and
              Nickson, Tom and
              Fujii, Keisuke. and
              {Boukouvalas}, Alexis and
              {Le{\'o}n-Villagr{\'a}}, Pablo and
              Ghahramani, Zoubin and
              Hensman, James},
  title        = "{{GP}flow: A {G}aussian process library using {T}ensor{F}low}",
  journal      = {Journal of Machine Learning Research},
  volume       = {18},
  number       = {40},
  pages        = {1-6},
  year         = {2017},
  month        = {apr},
  url          = {http://jmlr.org/papers/v18/16-537.html}
}

@article{Pereira2021,
author       = {Pereira, JB and Maciel, GF},
  title        = {Automated slump test: An effective alternative in predicting rheological properties and an efficient tool for providing the quality control of materials},
  journal      = {Measurement},
  volume       = {178},
  pages        = {109384},
  year         = {2021},
  publisher    = {Elsevier}
}

@article{Sao2021,
author       = {S{\'a}o, Yuri Taglieri and Pereira, Jo{\~a}o Batista and Fiorot, Guilherme Henrique and Maciel, Geraldo de Freitas},
  title        = {Obtaining flow curve for viscoplastic fluids through inclined open-channel apparatus},
  journal      = {Journal of the Brazilian Society of Mechanical Sciences and Engineering},
  volume       = {43},
  number       = {5},
  pages        = {278},
  year         = {2021},
  publisher    = {Springer}
}

@article{Valette2021,
author       = {Valette, Rudy and Pereira, A and Riber, S and Sardo, Lucas and Larcher, Aur{\'e}lien and Hachem, Elie},
  title        = {Viscoplastic dam-breaks},
  journal      = {Journal of Non-{N}ewtonian Fluid Mechanics},
  volume       = {287},
  pages        = {104447},
  year         = {2021},
  publisher    = {Elsevier}
}

@article{Bird1983,
author       = {Bird, R Byron and Dai, GC and Yarusso, Barbara J},
  title        = {The rheology and flow of viscoplastic materials},
  journal      = {Reviews in chemical engineering},
  volume       = {1},
  number       = {1},
  pages        = {1--70},
  year         = {1983},
  publisher    = {De Gruyter}
}

@book{Sarsby2000,
author       = {Sarsby, Robert W},
  title        = {Environmental geotechnics},
  year         = {2000},
  publisher    = {Thomas Telford}
}

@book{Blight2009,
author       = {Blight, Geoffrey E},
  title        = {Geotechnical engineering for mine waste storage facilities},
  year         = {2009},
  publisher    = {CRC press}
}

@article{Sosio2007,
author       = {Sosio, R and Crosta, GB and Frattini, P},
  title        = {Field observations, rheological testing and numerical modelling of a debris-flow event},
  journal      = {Earth Surface Processes and Landforms},
  volume       = {32},
  number       = {2},
  pages        = {290--306},
  year         = {2007},
  publisher    = {Wiley Online Library}
}

@article{Major1992,
author       = {Major, Jon J and Pierson, Thomas C},
  title        = {Debris flow rheology: Experimental analysis of fine-grained slurries},
  journal      = {Water resources research},
  volume       = {28},
  number       = {3},
  pages        = {841--857},
  year         = {1992},
  publisher    = {Wiley Online Library}
}

@article{OBrien1988,
author       = {O'Brien, Jim S and Julien, Pierre Y},
  title        = {Laboratory analysis of mudflow properties},
  journal      = {Journal of hydraulic engineering},
  volume       = {114},
  number       = {8},
  pages        = {877--887},
  year         = {1988},
  publisher    = {American Society of Civil Engineers}
}

@article{Ancey2007,
author       = {Ancey, Christophe},
  title        = {Plasticity and geophysical flows: A review},
  journal      = {Journal of non-{N}ewtonian fluid mechanics},
  volume       = {142},
  number       = {1-3},
  pages        = {4--35},
  year         = {2007},
  publisher    = {Elsevier}
}

@article{Gopala2008,
author       = {Gopala, Vinay R and Van Wachem, Berend Gm},
  title        = {Volume of fluid methods for immiscible-fluid and free-surface flows},
  journal      = {Chemical Engineering Journal},
  volume       = {141},
  number       = {1-3},
  pages        = {204--221},
  year         = {2008},
  publisher    = {Elsevier}
}

@article{Clayton2003,
author       = {Clayton, S and Grice, TG and Boger, DV},
  title        = {Analysis of the slump test for on-site yield stress measurement of mineral suspensions},
  journal      = {International journal of mineral processing},
  volume       = {70},
  number       = {1-4},
  pages        = {3--21},
  year         = {2003},
  publisher    = {Elsevier}
}

@article{Balmforth2014,
author       = {Balmforth, Neil J and Frigaard, Ian A and Ovarlez, Guillaume},
  title        = {Yielding to stress: Recent developments in viscoplastic fluid mechanics},
  journal      = {Annual review of fluid mechanics},
  volume       = {46},
  number       = {1},
  pages        = {121--146},
  year         = {2014},
  publisher    = {Annual Reviews}
}

@article{Cunha2014,
title={Uncertainty quantification through the Monte Carlo method in a cloud computing setting},
  author={Cunha Jr, Americo and Nasser, Rafael and Sampaio, Rubens and Lopes, H{\'e}lio and Breitman, Karin},
  journal={Computer Physics Communications},
  volume={185},
  number={5},
  pages={1355--1363},
  year={2014},
  publisher={Elsevier}
}

@article{Frigaard2005,
author    = {Frigaard, I. A. and Nouar, Cherif},
  title     = {On the usage of viscosity regularisation methods for visco-plastic fluid flow computation},
  journal   = {Journal of Non-Newtonian Fluid Mechanics},
  year      = {2005},
  volume    = {127},
  number    = {1},
  pages     = {1--26},
  doi       = {10.1016/j.jnnfm.2005.01.002},
}

@article{Tan2017,
author       = {Tan, Zhijun and Bernal, Susan A and Provis, John L},
  title        = {Reproducible mini-slump test procedure for measuring the yield stress of cementitious pastes},
  journal      = {Materials and Structures},
  volume       = {50},
  number       = {6},
  pages        = {235},
  year         = {2017},
  publisher    = {Springer}
}

@article{Bouvet2010,
author       = {Bouvet, Adrien and Ghorbel, Elhem and Bennacer, Rachid},
  title        = {The mini-conical slump flow test: Analysis and numerical study},
  journal      = {Cement and concrete research},
  volume       = {40},
  number       = {10},
  pages        = {1517--1523},
  year         = {2010},
  publisher    = {Elsevier}
}

@article{Jeyapalan1983,
author       = {Jeyapalan, Jey K and Duncan, J Michael and Seed, H Bolton},
  title        = {Analyses of flow failures of mine tailings dams},
  journal      = {Journal of Geotechnical Engineering},
  volume       = {109},
  number       = {2},
  pages        = {150--171},
  year         = {1983},
  publisher    = {American Society of Civil Engineers}
}

@article{Ancey2012,
author       = {Ancey, C and Andreini, N and Epely-Chauvin, G},
  title        = {Viscoplastic dambreak waves: Review of simple computational approaches and comparison with experiments},
  journal      = {Advances in Water Resources},
  volume       = {48},
  pages        = {79--91},
  year         = {2012},
  publisher    = {Elsevier}
}

@article{Muchiri2024a,
author       = {Muchiri, David Kibe and Hewett, James N and Sellier, Mathieu and Moyers-Gonzalez, Miguel and Monnier, Jerome},
  title        = {Numerical simulations of dam-break flows of viscoplastic fluids via shallow water equations},
  journal      = {Theoretical and Computational Fluid Dynamics},
  volume       = {38},
  number       = {4},
  pages        = {557--581},
  year         = {2024},
  publisher    = {Springer}
}

@article{Li2018,
author       = {Li, Xingyue and Zhao, Jidong},
  title        = {Dam-break of mixtures consisting of non-Newtonian liquids and granular particles},
  journal      = {Powder Technology},
  volume       = {338},
  pages        = {493--505},
  year         = {2018},
  publisher    = {Elsevier}
}

@article{Muchiri2024b,
author       = {Muchiri, David K and Cusack, Dale R and Hewett, James N and Sellier, Mathieu and Moyers-Gonzalez, Miguel and Monnier, Jerome},
  title        = {Identification of rheological parameters for shallow water flows of viscoplastic fluids using elevation hydrographs},
  journal      = {Physics of Fluids},
  volume       = {36},
  number       = {1},
  year         = {2024},
  publisher    = {AIP Publishing}
}

@article{BrondaniMinussi2025,
author       = {Brondani Minussi, Roberta and Canhoto Alves, Marcus Vin{\'\i}cius and de Freitas Maciel, Geraldo},
  title        = {Numerical Study on the Shear Stress Field Development on Dam Break Flows of Viscoplastic Fluids},
  journal      = {Fluids},
  volume       = {10},
  number       = {7},
  pages        = {180},
  year         = {2025},
  publisher    = {MDPI}
}

@article{Chaparian2018,
author       = {Chaparian, Emad and Nasouri, Babak},
  title        = {L-box—A tool for measuring the “yield stress”: A theoretical study},
  journal      = {Physics of Fluids},
  volume       = {30},
  number       = {8},
  year         = {2018},
  publisher    = {AIP Publishing}
}

@book{Coussot2005,
author       = {Coussot, Philippe},
  title        = {Rheometry of pastes, suspensions, and granular materials: applications in industry and environment},
  year         = {2005},
  publisher    = {John Wiley \& Sons}
}

@article{Coussot2014,
author       = {Coussot, Ph},
  title        = {Yield stress fluid flows: A review of experimental data},
  journal      = {Journal of Non-Newtonian Fluid Mechanics},
  volume       = {211},
  pages        = {31--49},
  year         = {2014},
  publisher    = {Elsevier}
}

@article{Perona2005,
author       = {Perona, Paolo},
  title        = {Bostwick degree and rheological properties: an up-to-date viewpoint},
  journal      = {Applied rheology},
  volume       = {15},
  number       = {4},
  pages        = {218--229},
  year         = {2005},
  publisher    = {Sciendo}
}

@article{Pereira2024,
author       = {Pereira, Jo{\~a}o Batista and Maciel, Geraldo de Freitas},
  title        = {Correlation between the automated slump test and the rheological properties of Portland cement mortars},
  journal      = {Magazine of Concrete Research},
  volume       = {76},
  number       = {12},
  pages        = {646--663},
  year         = {2024},
  publisher    = {Emerald Publishing Limited}
}

@article{Ferraris1998,
author       = {Ferraris, Chiara F and de Larrard, Fran{\c{c}}ois},
  title        = {Modified slump test to measure rheological parameters of fresh concrete},
  journal      = {Cement, Concrete, and Aggregates},
  volume       = {20},
  number       = {2},
  pages        = {241--247},
  year         = {1998},
  publisher    = {ASTM International}
}

@article{Choi2016,
author       = {Choi, Myoung Sung and Lee, Jung Soo and Ryu, Keum Seong and Koh, Kyung-Taek and Kwon, Seung Hee},
  title        = {Estimation of rheological properties of UHPC using mini slump test},
  journal      = {Construction and Building Materials},
  volume       = {106},
  pages        = {632--639},
  year         = {2016},
  publisher    = {Elsevier}
}

@article{Chidiac2000,
author       = {Chidiac, S. E. and Maadani, O. and Razaqpur, A. G. and Mailvaganam, N. P.},
  title        = {Controlling the quality of fresh concrete—a new approach},
  journal      = {Magazine of Concrete Research},
  volume       = {52},
  number       = {5},
  pages        = {353--363},
  year         = {2000}
}

@article{Zheng2024,
author       = {Zheng, Binbin and Zhao, Xiage and Wang, Jiahe and Wang, Wensong and Feng, Tingting and Yang, Yonghao},
  title        = {Analysis of the slump test for on-site yield stress measurement of thickened tailings},
  journal      = {Case Studies in Construction Materials},
  volume       = {21},
  pages        = {e03885},
  year         = {2024},
  publisher    = {Elsevier}
}

@article{Coussot1995,
author       = {Coussot, Philippe and Boyer, St{\'e}phane},
  title        = {Determination of yield stress fluid behaviour from inclined plane test},
  journal      = {Rheologica acta},
  volume       = {34},
  number       = {6},
  pages        = {534--543},
  year         = {1995},
  publisher    = {Springer}
}

@article{Lajeunesse2004,
author       = {Lajeunesse, E and Mangeney-Castelnau, A and Vilotte, JP},
  title        = {Spreading of a granular mass on a horizontal plane},
  journal      = {Physics of fluids},
  volume       = {16},
  number       = {7},
  pages        = {2371--2381},
  year         = {2004},
  publisher    = {American Institute of Physics}
}

@article{Nguyen2006,
author       = {Nguyen, TLH and Roussel, N and Coussot, P},
  title        = {Correlation between L-box test and rheological parameters of a homogeneous yield stress fluid},
  journal      = {Cement and Concrete Research},
  volume       = {36},
  number       = {10},
  pages        = {1789--1796},
  year         = {2006},
  publisher    = {Elsevier}
}

@article{Roussel2005b,
author       = {Roussel, Nicolas and St{\'e}fani, Christian and Leroy, Robert},
  title        = {From mini-cone test to Abrams cone test: measurement of cement-based materials yield stress using slump tests},
  journal      = {Cement and concrete research},
  volume       = {35},
  number       = {5},
  pages        = {817--822},
  year         = {2005},
  publisher    = {Elsevier}
}

@article{Saak2004,
author       = {Saak, Aaron W and Jennings, Hamlin M and Shah, Surendra P},
  title        = {A generalized approach for the determination of yield stress by slump and slump flow},
  journal      = {Cement and concrete research},
  volume       = {34},
  number       = {3},
  pages        = {363--371},
  year         = {2004},
  publisher    = {Elsevier}
}

@article{DellaVecchia2019,
author       = {Della Vecchia, Gabriele and Cremonesi, Massimiliano and Pisan{\`o}, Federico},
  title        = {On the rheological characterisation of liquefied sands through the dam-breaking test},
  journal      = {International Journal for Numerical and Analytical Methods in Geomechanics},
  volume       = {43},
  number       = {7},
  pages        = {1410--1425},
  year         = {2019},
  publisher    = {Wiley Online Library}
}

@article{Birnbaum2021,
author       = {Birnbaum, Janine and Lev, Einat and Llewellin, Edward W},
  title        = {Rheology of three-phase suspensions determined via dam-break experiments},
  journal      = {Proceedings of the Royal Society A},
  volume       = {477},
  number       = {2254},
  pages        = {20210394},
  year         = {2021},
  publisher    = {The Royal Society}
}

@article{Guo2024,
author       = {Guo, Junwei and Zhou, Qi and Wong, Ron Chik-Kwong},
  title        = {Determining the bulk viscosity in clay--sand mixtures: A combined investigation with slump tests and lattice Boltzmann simulations},
  journal      = {Physics of Fluids},
  volume       = {36},
  number       = {2},
  year         = {2024},
  publisher    = {AIP Publishing}
}

@article{Castruccio2010,
author       = {Castruccio, Angelo and Rust, AC and Sparks, RSJ},
  title        = {Rheology and flow of crystal-bearing lavas: Insights from analogue gravity currents},
  journal      = {Earth and Planetary Science Letters},
  volume       = {297},
  number       = {3-4},
  pages        = {471--480},
  year         = {2010},
  publisher    = {Elsevier}
}

@article{Zhang2024,
author       = {Zhang, Lianfu and Wu, Aixiang and Yang, Ke and Xia, Wentao and Zhang, Xi and Chen, Gezhong and He, Xiang},
  title        = {Application of lubrication theory to determine yield stress in slump tests},
  journal      = {Powder Technology},
  volume       = {444},
  pages        = {120081},
  year         = {2024},
  publisher    = {Elsevier}
}

@article{Gildeh2021,
author       = {Gildeh, Hossein Kheirkhah and Halliday, Alexandra and Arenas, Alfredo and Zhang, Hua},
  title        = {Tailings dam breach analysis: A review of methods, practices, and uncertainties},
  journal      = {Mine Water and the Environment},
  volume       = {40},
  number       = {1},
  pages        = {128--150},
  year         = {2021},
  publisher    = {Springer Nature BV}
}

@phdthesis{Machado2017,
author       = {Machado, Nath{\'a}lia Couto},
  title        = {Retroan{\'a}lise da propaga{\c{c}}{\~a}o decorrente da ruptura da barragem do fund{\~a}o com diferentes modelos num{\'e}ricos e hip{\'o}teses de simula{\c{c}}{\~a}o},
  school       = {Universidade Federal de Minas Gerais},
  year         = {2017}
}

@article{Saramito2017,
author    = {Saramito, Pierre and Wachs, Anthony},
  title     = {Progress in numerical simulation of yield stress fluid flows},
  journal   = {Rheologica Acta},
  year      = {2017},
  volume    = {56},
  number    = {3},
  pages     = {211--230},
  doi       = {10.1007/s00397-016-0992-1},
}

@phdthesis{Dias2017,
author       = {Dias, Neemias Almeida},
  title        = {Determina{\c{c}}{\~a}o de propriedades reol{\'o}gicas de rejeito de minera{\c{c}}{\~a}o por meio de re{\^o}metro rotacional},
  school       = {Universidade Federal do Espírito Santo},
  year         = {2017}
}

@phdthesis{Fitton2007,
author       = {Fitton, Timothy},
  title        = {Tailings beach slope prediction},
  school       = {RMIT University},
  year         = {2007}
}

@article{ElGarroussi2022,
author       = {El Garroussi, Siham and Ricci, Sophie and De Lozzo, Matthias and Goutal, Nicole and Lucor, Didier},
  title        = {Tackling random fields non-linearities with unsupervised clustering of polynomial chaos expansion in latent space: application to global sensitivity analysis of river flooding},
  journal      = {Stochastic Environmental Research and Risk Assessment},
  volume       = {36},
  number       = {3},
  pages        = {693--718},
  year         = {2022},
  publisher    = {Springer}
}

@book{Coussot2017,
author       = {Coussot, Philippe},
  title        = {Mudflow rheology and dynamics},
  year         = {2017},
  publisher    = {Routledge}
}

@book{Matsson2022,
  title={An introduction to ANSYS fluent 2022},
  author={Matsson, John E},
  year={2022},
  publisher={Sdc Publications}
}

@article{Jia2013,
  title={Kriging metamodeling for approximation of high-dimensional wave and surge responses in real-time storm/hurricane risk assessment},
  author={Jia, Gaofeng and Taflanidis, Alexandros A},
  journal={Computer Methods in Applied Mechanics and Engineering},
  volume={261},
  pages={24--38},
  year={2013},
  publisher={Elsevier}
}

@article{Lamboni2009,
title={Multivariate global sensitivity analysis for dynamic crop models},
  author={Lamboni, Matieyendou and Makowski, David and Lehuger, Simon and Gabrielle, Benoit and Monod, Herv{\'e}},
  journal={Field Crops Research},
  volume={113},
  number={3},
  pages={312--320},
  year={2009},
  publisher={Elsevier}
}
\end{document}